\documentclass[aps,prl,reprint,superscriptaddress,nofootinbib]{revtex4-2}

\usepackage{xcolor}

\usepackage{amsmath,amssymb}
\usepackage{graphicx}
\usepackage{bm}

\newcommand{\SM}{Supplemental Material}

\usepackage{mathtools}
\usepackage{pgfplots}
\pgfplotsset{compat=1.16}
\usetikzlibrary{calc,arrows.meta}
\definecolor{cB0}{RGB}{0,200,220}
\definecolor{cB1}{RGB}{120,120,120}
\definecolor{cB2}{RGB}{150,90,40}
\allowdisplaybreaks[3]
\newcommand{\dbimeas}{%
 \int_{S^2}\! d\theta_2 d\psi\,e^{-\phi^{\rm IIA}} r_h e^{Z}}

\newcommand{\dbidet}{\sqrt{-\det g}}
\newcommand{\dbiev}{\big|_{Z=0\leftrightarrow r=r_h}}
\newcommand{\dbiden}{2\dbidet_{(S^1_t\times\mathbb R^3)\times\mathbb R_{\geq0}}}
\newcommand{\LFmain}{FIG.~3 of the Letter}

\makeatletter
\renewcommand\addcontentsline[3]{}
\makeatother
\begin{document}

\title{Magnetic-slope crossovers in the QCD bulk-to-shear viscosity ratio}

\author{Aalok Misra}
\affiliation{Department of Physics, Indian Institute of Technology Roorkee, Roorkee, India}
\author{Shivam S. Kushwah}
\affiliation{Department of Physics, Indian Institute of Technology Roorkee, Roorkee, India}
\author{Keshav Dasgupta}
\affiliation{Department of Physics, McGill University, Montreal, QC, Canada}
\author{Charles Gale}
\affiliation{Department of Physics, McGill University, Montreal, QC, Canada}
\author{Sangyong Jeon}
\affiliation{Department of Physics, McGill University, Montreal, QC, Canada}

\begin{abstract}
How a magnetic field alters dissipation in quark--gluon matter is almost
unconstrained from first principles, since real-time transport is not directly
accessible to lattice simulation. Using a top-down MQGP-inspired
$\mathcal M$/string-theory construction of a QCD-like thermal plasma, together with
a kinetic-theory identity within a specified single-mode quasiparticle truncation
and a compatibility construction using available lattice equation-of-state data,
we propose a normalization-insensitive sign prediction that the field slope of the
longitudinal bulk-to-shear viscosity ratio $\zeta_\parallel/\eta_\parallel$
reverses sign \emph{twice} with temperature---below and above the
deconfinement pseudo-critical temperature $T_0$, at $T_1<T_0<T_2$ (the lower
reversal $T_1$ conditional on a stated pion-gas rate inequality). A
kinetic-theory identity with phase-appropriate carriers, a quark--gluon plasma
above $T_0$ and a chiral pion gas below, reproduces the same pattern, and
lattice speed-of-sound data can be shown to be consistent as well. The bottom-up VQCD in the deconfined phase supports a negative slope for large temperatures; the slope crossover is not resolvable from the available $\frac{\zeta_\parallel}{s}, \frac{\eta_\parallel}{s}$ plots. 
\end{abstract}

\maketitle

\paragraph{\bf Introduction.}
A magnetic field is one of the few external handles that can be applied to
strongly interacting matter both in the laboratory and in the early universe.
Non-central heavy-ion collisions generate the strongest fields known,
transiently of order $eB\sim m_\pi^2$; magnetars sustain comparable fields for
astrophysical times. What such a field does to the \emph{equilibrium}
properties of QCD matter---the equation of state, the chiral condensate, the
transition temperature---is by now well mapped by lattice simulation. What it
does to \emph{dissipation} is not.

The obstruction is structural rather than technical. Viscosities are real-time
response coefficients, defined by Kubo formulae involving the analytic
continuation of Euclidean correlators; extracting them from lattice data
requires an ill-posed inversion, and doing so in a background field has not
been achieved. Transport in magnetized QCD is therefore a domain where a
normalization-insensitive \emph{sign} prediction---as opposed to a fit or a
model---has real value, because there is currently nothing to fit to. We are
explicit about the logical hierarchy: the top-down construction \emph{proposes}
the sign pattern; the kinetic identity \emph{explains} the sign mechanism within
a controlled single-mode truncation; the lattice equation of state
\emph{demonstrates compatibility} (not a determination of $\zeta/\eta$); and
VQCD \emph{checks} the high-$T$ negative-slope trend only.

We make such a prediction here. The quantity we track is the ratio of bulk-to-shear viscosity, measured along the field direction. Bulk viscosity $\zeta$
measures resistance to uniform compression and vanishes for a scale-invariant
fluid; shear viscosity $\eta$ measures resistance to shape deformation. Their
ratio is a clean diagnostic of how far QCD sits from conformality.

Our claim is about the \emph{slope} of this ratio in the field, at fixed
temperature. We find that
\begin{equation}
\frac{\partial}{\partial B}\Big(\frac{\zeta_\parallel}{\eta_\parallel}\Big)\bigg|_T
\end{equation}
changes sign \emph{twice} as $T$ is raised through the transition region:
it is positive at low $T$, turns negative at a temperature above $T_1<T_0$ ($T_0$ being the reduced transition temperature in the presence of a weak magnetic field) still inside
the confined phase, and---after the deconfinement transition at $T_0$---turns
positive/negative at temperatures below/above $T_2>T_0$ in the deconfined phase. With standard
inputs, $T_1\simeq134$~MeV, $T_0\simeq143$~MeV and $T_2\simeq187$~MeV.

Two features make this worth stating as a Letter. First, it is a
\emph{qualitative} statement---a pattern of signs---and therefore insensitive
to the overall normalizations  of any holographic
computation. Second, it is reached from two directions with no common dynamical input ---a
top-down $\mathcal M$-theory dual and a kinetic-theory identity -- while lattice
equation-of-state data can be made \emph{consistent with} both (a compatibility
construction, not an independent determination of $\zeta/\eta$). Interestingly, a bottom-up VQCD holographic model is shown also to be consistent in the sense of obtaining a negative large-temperature slope in the deconfined phase even though the crossover in the same phase is not graphically resolvable from the available $\frac{\zeta_\parallel}{s}, \frac{\eta_\parallel}{s}$ plots.

\vskip.1in

\paragraph{\bf What is being computed.}
In a magnetic field the rotational symmetry of the medium is reduced from
$SO(3)$ to $SO(2)$ about the field axis, and a single shear and a single bulk
viscosity no longer suffice. The full classification of the resulting
transport coefficients for a relativistic magnetized fluid is discussed in ~\cite{HK2017}: the shear sector splits into longitudinal
and transverse coefficients $\eta_\parallel,\eta_\perp$ together with two
non-dissipative Hall viscosities, and the bulk sector is governed by four
coefficients of which only three are independent, by an Onsager relation.

We track one specific combination in kinetic theory ---the longitudinal, parity-even projection
\begin{equation}
\mathcal R(B,T)\equiv\frac{\zeta_\parallel(B,T)}{\eta_\parallel(B,T)}
=\mathcal R_0(T)+\mathcal R_2(T)\Big(\frac{B}{T_c^2}\Big)^{\!2}+\cdots,
\label{eq:Rdef}
\end{equation}
obtained by projecting the viscosity tensors along the field direction. The
longitudinal projection returns a genuine Kubo coefficient rather than a bare
tensor label: the Eling--Oz horizon datum responds to $B$ only through the single
scalar $r_{h/0}(B)$, so the induced compression is isotropic at leading order and
the cross viscosity $\zeta_{\parallel\perp}$---which would enter through the
anisotropic ($\ell{=}2$) part of the horizon expansion---is $N$-suppressed beyond
$\mathcal O(1/N)$ by the same fractional-anisotropy estimate that controls the
projection anisotropy [\SM~\cite{SM}, Sec.~I]. Writing the expansion in the dimensionless variable
$B/T_c^2$ makes the comparison between frameworks unambiguous. In the equilibrium
background ($\Omega$(vorticity)=$E$(electric field)$=0$) the field $B$ is $P$-even and $T$-odd---it is the only
axial vector present---so a (pseudo)scalar(e.g. $(B.a)^{n_B}, a^\mu$ being the acceleration - a vector)'s parity is $(-1)^{n_B}$ and $\mathcal R$,
built from the even-in-$B$ (longitudinal magnetization and viscosity) structures
of the Hern\'andez--Kovtun classification, is parity-even; hence
$\partial_B\mathcal R|_{B=0}=0$ identically. By ``field slope''
we therefore mean $\partial_B\mathcal R$ evaluated at fixed small $B>0$, where
$\partial_B\mathcal R=2B\,\mathcal R_2(T)/T_c^4+O(B^3)$, so its sign is that of
the weak-field curvature coefficient $\mathcal R_2(T)$---the quantity that
actually changes sign, and the one the figures resolve at finite $B$. The ${\cal M}$/string-theory top-down dual instead
yields a series in $B/T_c^2$ in which a parity-odd linear-in-$B$ term
\emph{does} occur---admitted by the genuinely parity-odd type-IIB axion and
warped five-form background, not by $B$ itself, and carried on a separate
projection--- in analogy with \cite{Maldacena+Martelli}, referring to (S5) and the discussion following (S27) of the supplemental file, interestingly, the ratio of the ${\cal O}(B)$ and ${\cal O}(B^2)$ coefficients can be interpreted as the product of the cube roots respectively of the real and imaginary parts of the periods of the non-compact and compact three-cycles of a deformed conifold and is additionally large $N$-suppressed (with this ratio, numerically, working out to a little under $\approx1\%$); the lattice $c_s^2$-vs-$T$ data at several $B$s are shown to be
compatible with this structure (a parity-even quadratic term with a percent-level
parity-odd linear admixture) while ensuring a smooth crossover at $T_0$ from lattice EOS data; given that (for simplicity) we work in the $N_f=3$ chiral limit inn the string/${\cal M}$-theory dual, we obtain a first-order Hawking-Page phase transition therein (the double magnetic slope crossover is common to both though).  All definitions and the weak-field power counting, including the $(-1)^{n_B}$ parity rule and the retained even-in-$B$ structures, are given in Sec.~I of the \SM~\cite{SM}.

\vskip.1in

\paragraph{\bf A kinetic-theory identity.}
The organizing tool is an identity that is exact \emph{within the single-mode
variational quasi-particle truncation}, where the bulk and shear channels share
the same scalar weight, so common phase-space and Landau-level factors cancel in
the ratio (this cancelation is a property of the truncation, not a theorem of
magnetized QCD transport) for how the ratio responds to the field. Writing $X=\zeta,\eta$ in the standard variational form $X=\mathcal N_X^2/\Gamma_X$---a source
projection $\mathcal N_X$ squared over a collision rate $\Gamma_X$---and
differentiating the logarithm of the ratio gives
\begin{equation}
\frac{\partial\ln(\zeta/\eta)}{\partial(q_fB)^2}
=2\big(\nu_\zeta-\nu_\eta\big)-\big(\gamma_\zeta-\gamma_\eta\big),
\label{eq:master}
\end{equation}
where $\nu_X$ and $\gamma_X$ are the field log-derivatives of the source and
of the rate, respectively. Equation~\eqref{eq:master} is an identity that holds in both phases, and all the Physics is in
which quasi-particles supply $\mathcal N_X$ and $\Gamma_X$.

Its usefulness is that the two terms have transparent and \emph{competing}
physical content. The source difference is controlled  by the
conformal-breaking factor $\tfrac13-c_s^2$, since only the bulk source carries
it:
\begin{equation}
2\big(\nu_\zeta-\nu_\eta\big)=-\frac{4\chi_B}{\tfrac13-c_s^2},
\qquad
\chi_B\equiv\frac{\partial c_s^2}{\partial (q_fB)^2}.
\label{eq:source-piece}
\end{equation}
Because $\tfrac13-c_s^2$ is small, it is the \emph{driver} of the effect. The
rate difference $\gamma_\zeta-\gamma_\eta$ acts as the competitor, and the
sign of $\partial_B\mathcal R$ is decided by which of the two wins. A crossover
occurs wherever the balance tips. Derivations of \eqref{eq:master} and
\eqref{eq:source-piece}, including the cancellations that make the
common phase-space factors drop out of the difference, are in Sec.~II of the
\SM~\cite{SM}.

\vskip.1in

\paragraph{\bf The top-down input.}
The top-down input comes from a UV-complete type IIB string/type IIA mirror/$\mathcal M$-theory uplift
 of a QCD-like thermal plasma in the MQGP
limit~\cite{FEP,MQGP,Kushwah2024} --- a holographic model with (large) $(M)N$ color (fractional) $D3$-branes, finite
intermediate coupling and wrapped flavor $D7$-branes designed to reproduce several realistic QCD-like observables in appropriate regimes in which the
magnetic field appears as a supergravity field localized near the black-brane horizon and the flavor
sector is realized by wrapped branes. The dual has been checked against lattice and PDG data across glueball \cite{MQGP-glueball-spectroscopy} and
meson spectra \cite{MQGP-meson-spectroscopy}, the trace anomaly \cite{MQGP-trace-anomaly} and several transport coefficients \cite{IITR-McGill-Bulk}, \cite{MQGP-transport}, \cite{Kushwah2024}. So it has not been (fine) tuned to $\zeta$.

For the viscosity ratio the relevant statement is due to Eling and Oz \cite{ElingOz2011}: at the
two-derivative level $\zeta/\eta$ is fixed by the rate at which the internal
moduli of the geometry flow along the family of black-brane solutions labeled
by temperature, evaluated at the horizon. Because both viscosities are horizon
responses with the same normalization, the ratio is free of the overall
factors that plague individual transport coefficients---the reason we
predict a ratio rather than $\zeta$ or $\eta$ separately.

A concern specific to a UV-complete construction is that the dual contains
higher-derivative $O(R^4)$ corrections, which could in principle modify the
Eling-Oz (EO) prescription itself rather than merely deform the background. We
show that they do not, at the order worked: the higher-derivative corrections
to the horizon entropy (Wald)  are suppressed by an
additional power of $1/N$ relative to the terms retained, so evaluating the
isotropic two-derivative formula on the ${\cal O}(R^4)$-corrected background is the
consistent truncation rather than an assumption. A projection ($\parallel$
vs.\ $\perp$) anisotropy of the ratio would arise only from an anisotropic bulk
metric; here the geometry is \emph{isotropic}, the field being a
constant on the flavor $D6$ world-volume whose only leading backreaction is a
shift of the single scalar horizon datum $r_{h/0}(B)$ (the horizon radius for
$T>T_c$, the IR cutoff for $T<T_c$). Since $\zeta$ and $\eta$ depend on $B$ only
through $r_{h/0}(B)$, the ratio is a function of that one scalar. The residual
metric anisotropy is bounded by its fractional size: solving the linearized
color Einstein equation with the anisotropic ($\ell{=}2$) $D6$-brane energy-momentum tensor,
$\delta g_{x^3x^3}/g_{x^3x^3}=(\Delta_L^{-1}T^{D6,\ell=2})_{x^3x^3}/({\cal G}^{-1}
T^{\rm SUGRA})_{x^3x^3}$, where $\Delta_L$ is the Lichnerowicz operator (the leading part
of the linearized color Einstein operator ${\cal G}$ on the transverse-traceless
$\ell{=}2$ sector). Using the explicit MQGP scalings this is evaluated in \cite{SM}'s SIII.I to yield
$\delta g_{x^3x^3}/g_{x^3x^3}\sim (g_s^{1/2}N_f/N^{7/4})\,B^2$, i.e.\
${\cal O}(1/N^{1+x_g})$ with $x_g>0$: the metric anisotropy, though of the same
$B^2$ order as $\mathcal R$, is $N$-suppressed and may be dropped. The residual
world-volume flux, being $SO(2)$-averaged, feeds only the isotropic coefficient
and induces no $\parallel/\perp$ difference. The sign of
$\mathcal R_2$ that the claim uses is therefore unaffected. This is the technical
result that licenses the calculation; it is
derived in Sec.~III of the
\SM~\cite{SM}, together with the explicit brane-worldvolume computation. Further, the EO prescription sets $\frac{\eta}{s}$ at its conformal value. As it was shown in \cite{IITR-McGill-Bulk} that this ratio in our framework receives non-conformal corrections that are $\frac{1}{N}$-suppressed, and that $\frac{\zeta}{s}$ is also $\frac{1}{N}$-suppressed, it suffices to set, like the EO assumption, $\frac{\eta}{s}$ at its conformal value.

The outcome is that the top-down dual returns $\partial_B\mathcal R>0$ at low
temperature and $\partial_B\mathcal R<0$ at high temperature, on either side of $T_0$.

\vskip.1in

\paragraph{\bf Why the slope must reverse.}
The mechanism can be explained from kinetic theory. Both reversals follow from
the competition in Eq.~\eqref{eq:master} between the amplified source term and
the collision rate term, with the microscopic input differing between phases.

\begin{figure}[t]
\centering
\includegraphics[width=0.95\columnwidth]{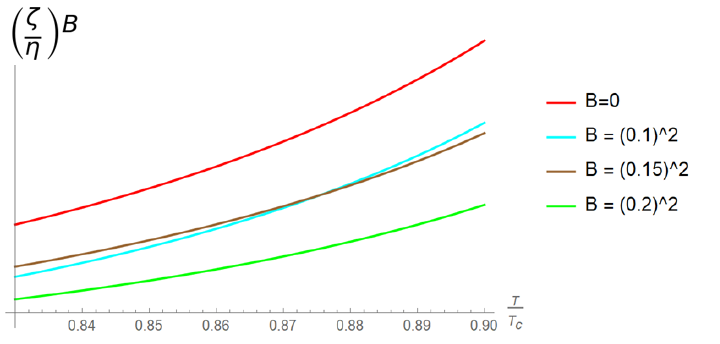}
\caption{Confined phase ($T<T_0$): the top-down string prediction for
$\zeta/\eta$ versus $T/T_0$ at weak field. The $B{=}(0.1)^2$ and
$B{=}(0.15)^2$ curves cross near $T/T_0\simeq0.87$, so the field slope
$\partial_B(\zeta/\eta)|_T$ turns from positive to negative there. This is the
lower reversal $T_1$. The independent lattice construction of
Sec.~V of the \SM~\cite{SM} places the same reversal at $T_1\simeq134$~MeV
($T/T_0\simeq0.94$).}
\label{fig:confined}
\end{figure}

\emph{Below $T_0$}, the carriers are pions---the pseudo-Goldstone bosons of
chiral symmetry breaking. The field enters chiefly by raising the charged-pion
mass to its lowest-Landau-level value, $m^2(|eB|)=m^2+|eB|$. Two competing
effects again appear: the field-induced change in the speed of sound (inverse
magnetic catalysis, $\tilde{\chi}<0$, giving a positive contribution to the
slope near $T\ll m$) and the growth of the pion-gas collision rates as $T$
approaches $T_0$. Evaluating the endpoints in the chiral limit shows the slope is positive as
$T\ll m$ and may be negative as $T\to T_0^-$ if an inequality ((S28) of \cite{SM}'s SV) is satisfied, so a reversal at some $T_1$ inside the
confined phase follows; the top-down holographic curves realizing this reversal are shown
in Fig.~\ref{fig:confined}. We state explicitly the sufficient condition on the
pion-gas rates under which the endpoints straddle; the full magnetized
pion-gas transport calculation that would establish it from first principles
has not been performed, here or in \cite{SM}, and we flag $T_1$ accordingly. The
endpoint analysis, the straddle condition and the chiral-limit bookkeeping are
in Sec.~V of the \SM~\cite{SM}.

\begin{figure}[t]
\centering
\includegraphics[width=0.95\columnwidth]{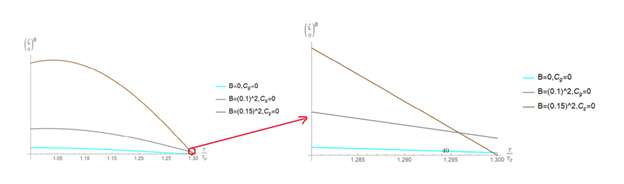}
\caption{Deconfined phase ($T>T_0$): the top-down prediction for
$\zeta/\eta$ versus $T/T_0$ at weak field
(left: $[T_0,1.3\,T_0]$; right: zoom on $[1.28,1.30]\,T_0$). The curves for
different $B$ converge and cross near the right edge, realising
$\partial_B(\zeta/\eta)|_T>0\to<0$ at the upper reversal
$T_2\simeq187$~MeV. Together with Fig.~\ref{fig:confined} this gives the
four-region sign pattern $\partial_B(\zeta/\eta)>0$ ($T<T_1$), $<0$
($T_1<T<T_0$), $>0$ ($T_0<T<T_2$), $<0$ ($T>T_2$), with $T_1<T_0<T_2$ ---
the falsifiable content of this Letter \cite{Cg}.}
\label{fig:crossover}
\end{figure}

\emph{Above $T_0$}, the carriers are quarks and gluons. The field enters
through the magnetized quark loop, whose contribution to the rates is
controlled by a thermal-mass parameter $\alpha\sim m_{\rm loop}/T$. At  $T_0<T<m_{\rm loop}$, the contribution from the source term is
exponentially suppressed, leaving the collision term with a positive
slope. At $T>m_{\rm loop}$, the source ($\eta_\zeta-\eta_\eta$) and collision $-(\gamma_\zeta-\gamma_\eta)$ terms' contributions are both negative, giving a negative slope. Since the two limits have opposite signs
and everything is continuous, a zero must lie between them: the crossover
$T_2$ is a \emph{consequence} of the two limits, not an input. The perturbative
quasiparticle estimate supplies only the \emph{scale and existence} of $T_2$,
not a controlled value---near $150$--$200$~MeV weak coupling is not under
control; the quoted $T_2\simeq187$~MeV is read from the lattice construction
($\xi_2{=}0$) consistent with the the top-down holographic curves
realizing this reversal are shown in Figs.~\ref{fig:confined} and\ref{fig:crossover}.

\vskip.1in

\begin{figure*}[tp]
\centering
\includegraphics[width=0.72\textwidth,height=0.86\textheight,keepaspectratio]{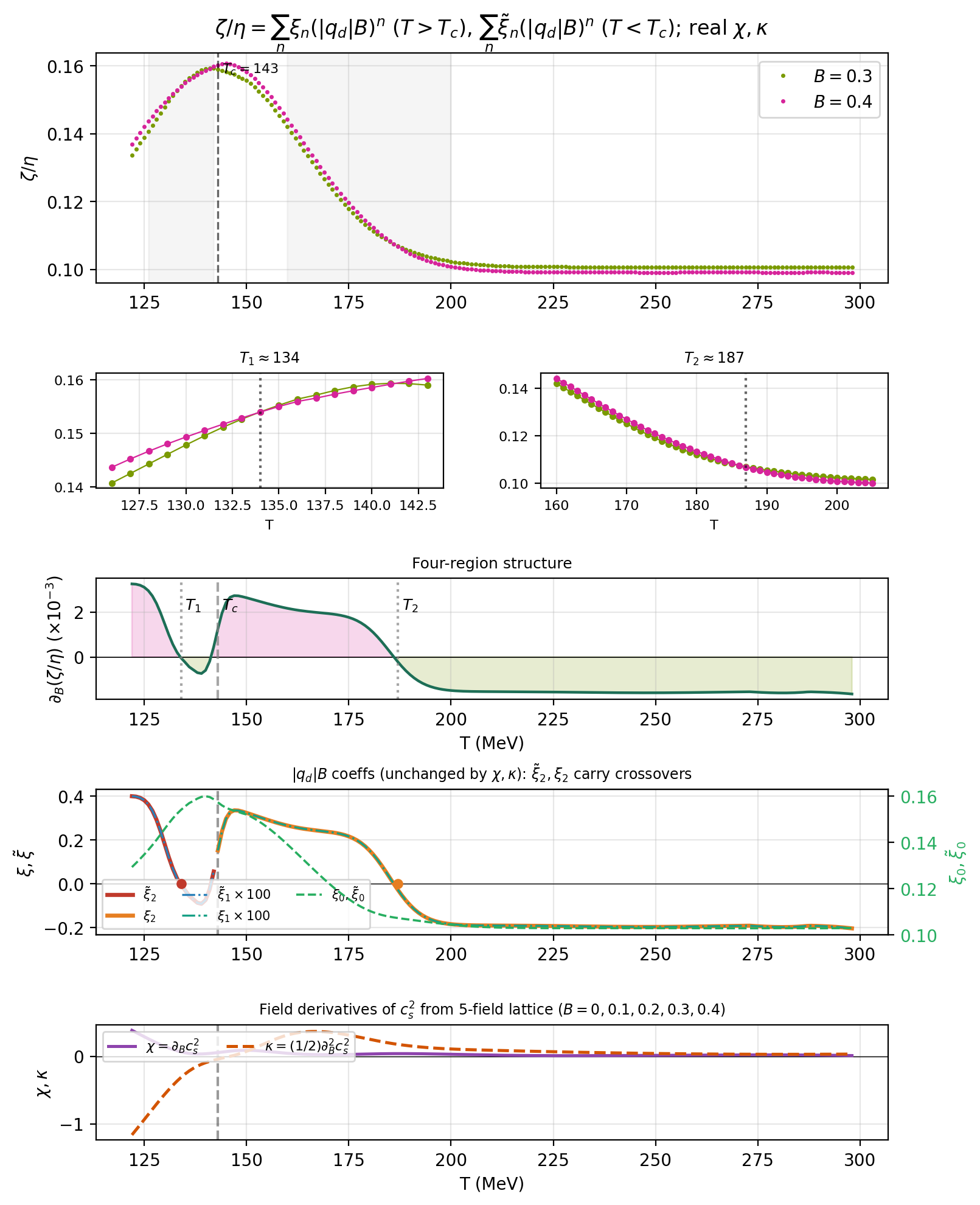}
\caption{$\zeta/\eta$ vs $T$ for $B=0.3,0.4$ in the natural field variable
$|q_d|B=B/3$, from the lattice construction of Sec.~V of the
\SM~\cite{SM} for physical $N_f=2{+}1$ quark masses
($\widetilde\xi_n$ for $T<T_0$, $\xi_n$ for $T>T_0$), continuous at
$T_0\simeq143$~MeV as required for an analytic crossover. These curves are a
\emph{model construction} built on lattice $c_s^2(B,T)$ input, not a direct
lattice determination of $\zeta/\eta$.
\textit{Top:} full curves through $T_0$ (dashed).
\textit{Middle:} zooms near the lower crossover $T_1\simeq134$~MeV and the
upper crossover $T_2\simeq187$~MeV.
\textit{Third:} the $\partial_B(\zeta/\eta)\times10^{-3}$, \emph{continuous at
$T_0$}, generating the four-region sign pattern (positive below $T_1$,
negative on $(T_1,T_0)$, positive on $(T_0,T_2)$, negative above $T_2$)
with $T_1<T_0<T_2$.
\textit{Fourth:} the $|q_d|B$-expansion coefficients; $\widetilde\xi_2,\xi_2$
vanish at $T_1,T_2$ and the parity-odd linear-in-$B$ coefficients
$\widetilde\xi_1,\xi_1$ ($\times100$) sit at $\sim1\%$ of the quadratic ones.
\textit{Bottom:} the field derivatives $\chi=\partial_Bc_s^2$ and
$\kappa=\tfrac12\partial_B^2c_s^2$ from the five-field lattice set.
Crossover temperatures carry a few-MeV uncertainty from the $c_s^2$
digitisation.}
\label{fig:fig4}
\end{figure*}

\paragraph{\bf Consistency with lattice data.}
Although the viscosities themselves are not available from the lattice, the
quantity that drives the source term in Eq.~\eqref{eq:source-piece}---the
field dependence of the speed of sound via the EoS ---is. Using published 
at several field strengths, one can construct the field expansion of the ratio
phase by phase and inspect the resulting slope. The plots in Fig.~\ref{fig:fig4} utilize published EoS-data \cite{Bali14} (from which one can obtain $c_s^2$-vs-$T$ plots for different $B$s), and by design, are continuous across $T_0$, permit a small admixture of parity-violating terms linear in $B$, are predominantly given by $1/3 - c_s^2$ in the deep confined phase and $(1/3 - c_s^2)^2$ in the deep deconfined phase. They then naturally exhibit the double crossover, as obtained from ${\cal M}$/string theory and kinetic theory.

The lattice comparison  demonstrates that the predicted crossover structure can be made to be compatible with what is known about the magnetized lattice equation of state---it is not an independent determination of $\zeta/\eta$. The explicit construction, the coefficient extraction and the error budget are in Sec.~V of the \SM~\cite{SM}.

\begin{figure*}[t]
\includegraphics[width=\textwidth]{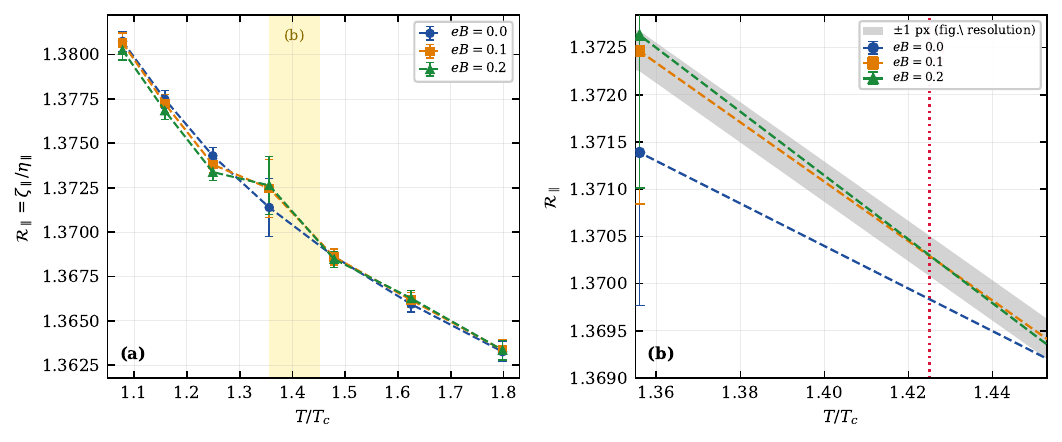}
\caption{\label{fig:vqcd}
$\mathcal R_\parallel=\zeta_\parallel/\eta_\parallel$ in bottom-up V-QCD,
digitized from FIG.~1 of \cite{Demircik2023L} using
$\eta_\perp/s=1/4\pi$, $\eta_\parallel/s=e^{-2W_H}/4\pi$ (Eq.~(16) of
\cite{Demircik2023L}, quoted there from \cite{Jain+Samanta+Trivedi},
arXiv:1506.01899). Here $W$ is the
metric anisotropy function in the five-dimensional Einstein-dilaton ansatz of
\cite{Demircik2023L},
$ds^2=e^{2A(r)}\big[f(r)^{-1}dr^2-f(r)dt^2+e^{2W(r)}(dx^2+dy^2)+dz^2\big]$,
with the magnetic field $\bm B=B\hat z$ realized by the bulk gauge field
$A_\mu=(0,-yB/2,xB/2,0,0)$, so that $e^{2W}$ measures the distortion of the
plane transverse to $\bm B$ relative to the field direction; $W_H\equiv
W(r_H)$ denotes its value at the horizon, and $W_H\to0$ as $B\to0$, recovering
the isotropic result $\eta_\parallel/s=\eta_\perp/s=1/4\pi$ (validated by
recovering $\eta_\parallel/s(B{=}0)=1/4\pi$ to $<0.11\%$); error bars are
digitization $1\sigma$.
\textbf{(a)} Published range $1.08\le T/T_c\le1.80$; shading marks the region
enlarged in (b).
\textbf{(b)} Zoom on $1.355\le T/T_c\le1.453$, about $T=1.425\,T_c$, where the
$eB=0.1,0.2$ curves appear to cross.  Their separation ($\lesssim2$ px) lies
below the source figure's linewidth (grey band), and the bracketing points are
consistent with zero ($+0.07\sigma$, $-0.19\sigma$): the crossing is
\emph{not resolvable} from the published plot.}
\end{figure*}

A fourth line of evidence comes from V-QCD \cite{Demircik2024}, a bottom-up holographic model fitted to lattice thermodynamics. In its published form, it supports $\partial_B\mathcal R_\parallel<0$ at high temperatures in the deconfined phase; the available plotted data do not resolve a sign reversal - see FIG. \ref{fig:vqcd}. We hence treat it as a compatibility check rather than a fully independent confirmation.

\vskip.1in

\paragraph{\bf What would falsify this.}
The prediction is a sign pattern with two zeros, and it is falsifiable in
several ways. A determination of $\zeta/\eta(B,T)$---from a future lattice
computation of magnetized transport, from Bayesian extraction against
heavy-ion data with field-dependent transport coefficients---that found a monotonic field slope across the transition region, or zeros at substantially different
temperatures, would contradict it. Conversely, observation of the double
reversal would constitute evidence that the near-transition region is
described by carriers whose conformal-breaking response and collision rates
behave as sketched here.

The lower crossover $T_1$ rests on a
sufficiency condition on pion-gas rates that we have not derived; it should be
regarded as a prediction conditional on that condition. 

We make the logical structure explicit, since some intermediate top-down steps
(the detailed magnetic-field construction, the full DBI determinant, and the
free-energy matching) are given in companion works. The \emph{falsifiable}
content of this Letter---the four-region sign pattern with $T_1<T_0<T_2$---is
carried by the kinetic-theory identity Eq.~\eqref{eq:master}. The lattice $c_s^2(B,T)$,  fixes $\mathrm{sgn}\,\chi_B$ (that is proved to be naturally so in the top-down computation) and hence its sign structure without reference to the companion computations. The top-down $\mathcal M$-theory dual enters as an independent corroboration of the same sign, using only the horizon-datum sign.
The reader can thus verify the central claim from the kinetic and unconditionally from the top-down construction; the EOS lattice results can be made to show the aforementioned four-region sign pattern of the magnetic slope, with a smooth crossover at the transition temperature - the top-down computation yields a Hawking-Page phase transition.

\vskip.1in

\paragraph{\bf Summary.}
We have proposed that the magnetic-field slope of the longitudinal
bulk-to-shear viscosity ratio of QCD changes sign twice across the
deconfinement region, at $T_1<T_0<T_2$: the top-down $\mathcal M$-theory
construction \emph{proposes} the pattern, a kinetic-theory identity with
phase-appropriate carriers \emph{explains} the mechanism within a single-mode
truncation. Lattice equation-of-state data, imposing a smooth crossover at a transition temperature (unlike the discontinuity due to the first-order Hawking-Page phase transition obtained in the top-down computation) -- shifted to lower values in the presence of a magnetic field -- as well as the ratio approaching $1/3 - c_s^2$ deep into the strongly coupled confined phase and $(1/3 - c_s^2)^2$ deep into the weakly coupled deconfined phase, is shown to be \emph{compatible} with the aforementioned double field slope crossover (a compatibility construction, not a determination of $\zeta/\eta$). A bottom-up holographic VQCD model supports the negative slope at large temperatures in the deconfined phase; the field slope crossover is not resolvable from the available $\frac{\zeta_\parallel}{s}$ and $\frac{\eta_\parallel}{s}$. The robust claim is the \emph{existence} of the double reversal; the numerical locations (in particular $T_2\simeq187$~MeV) are the values produced by the lattice-EoS-compatible parametrization used here and should be regarded as indicative. It is, to our knowledge, the first
normalization-insensitive sign statement about the field dependence of a dissipative ratio
in QCD across the transition. All derivations, the ${\cal O}(R^4)$ analysis, the
brane computation, the pion-gas endpoints and the lattice construction are
given in the \SM~\cite{SM}.

\vskip.1in

\clearpage
\onecolumngrid
\begin{center}\rule{0.6\textwidth}{0.4pt}\end{center}
\twocolumngrid
\setcounter{secnumdepth}{3}
\setcounter{section}{0}
\setcounter{equation}{0}
\setcounter{figure}{0}
\setcounter{table}{0}
\renewcommand{\thesection}{S\Roman{section}}
\renewcommand{\thesubsection}{\thesection.\arabic{subsection}}
\renewcommand{\theequation}{S\arabic{equation}}
\renewcommand{\thefigure}{S\arabic{figure}}
\renewcommand{\thetable}{S\arabic{table}}

\begin{center}\textbf{\large Supplemental Material}\end{center}
\medskip

\noindent
This Supplemental Material collects the definitions, derivations and numerical
constructions underlying the Letter.  Section~\ref{sm:obs} fixes the
observable and the weak-field power counting; Sec.~\ref{sm:kin} derives the
kinetic-theory master identity and the sign of the rate terms;
Sec.~\ref{sm:top} gives the $\mathcal M$-theory construction, the ${\cal O}(R^4)$
analysis and the brane-worldvolume computation; Sec.~\ref{sm:cross}
establishes the two slope crossovers; Sec.~\ref{sm:pion} treats the
confined-phase pion gas and the endpoint/straddle analysis;
Sec.~\ref{sm:lat} gives the lattice construction across the crossover; and
Sec.~\ref{sm:mutual} discusses the mutual consistency of the four frameworks - top-down holography (prediction), kinetic theory (supports), lattice theory (available EoS lattice data can be made to be consistent with string/${\cal M}$-theory prediction) and bottom-up VQCD (match in the large-temperature limit in the deconfined phase; crossover feasible but graphically not resolveable). Equation, figure and table numbers carry an ``S'' prefix; references to
unprefixed equations refer to the Letter.


\section{The observable and weak-field power counting}
\label{sm:obs}

\subsection{Introduction}
\label{intro}
Transport coefficients of QCD matter in a magnetic field are poorly
constrained: lattice determinations of dissipative quantities such as
$\zeta$ and $\eta$ in a background field are scarce, since real-time
transport is not directly accessible to Euclidean simulation.  A
first-principles, top-down holographic statement about $\zeta/\eta(B,T)$
is therefore a genuine prediction rather than a fit.  We make one here:
the field derivative of $\zeta/\eta$ at fixed $T$ changes sign as $T$ is
varied, $\partial_B(\zeta/\eta)|_T>0\to<0$, with slope reversals on both sides
of the pseudo-critical temperature $T_0$.  We first fix the observable
precisely.  In the magnetized theory spatial $SO(3)$ breaks to $SO(2)$, so
the shear sector splits into $\{\eta_\parallel,\eta_\perp\}$ (plus Hall pieces
$\tilde\eta_\parallel,\tilde\eta_\perp$) and the bulk sector into three
independent coefficients, following the relativistic-magnetohydrodynamic
classification \cite{HK2017}; in the $\{\zeta_\parallel,\zeta_\perp,\zeta_\times\}$ nomenclature
\cite{Cao2022}, $\zeta_\times$ is the longitudinal--transverse cross bulk
viscosity coupling $\partial_z u_z$ to $\partial_\perp\!\cdot u_\perp$.  We
write the constitutive relation
$\delta T_{ij}=-\eta_{ijkl}\big(\partial_k u_l+\partial_l u_k
-\tfrac23\delta_{kl}\partial\!\cdot\!u\big)-\zeta_{ij}\,\partial\!\cdot\!u
+\cdots$, where $\eta_{ijkl}(B,T)$ is the rank-4 shear-viscosity tensor
(symmetric, traceless on $kl$) and $\zeta_{ij}(B,T)$ the rank-2 bulk-viscosity
tensor \cite{HK2017}, both reducing to $\eta(\delta_{ik}\delta_{jl}+\dots)$ and
$\zeta\,\delta_{ij}$ at $B{=}0$. In the equilibrium magnetized background the fluid is static,
$u^{\mu}=(1,\mathbf{0})$, and both the vorticity and the electric field vanish,
$\Omega^{\mu}=0$ and $E^{\mu}=0$ (the latter by electric screening: in the MHD
counting $B\sim\mathcal{O}(1)$ while $E\sim\mathcal{O}(\partial)$, so $E\to0$ in
strict equilibrium). Under spatial parity the magnetic field is an axial vector,
$B^{\mu}\to+B^{\mu}$, whereas the acceleration $a^{\mu}$, the velocity $u^{i}$,
and $\nabla_{i}$ are all polar. Once $\Omega$ and $E$ are set to zero, $B^{\mu}$
is therefore the \emph{only} axial vector in the problem, and the parity of any
scalar is fixed simply by counting factors of $B$,
\begin{equation}
P\;=\;(-1)^{n_{B}},
\end{equation}
so that linear-in-$B$ structures are $P$-odd and $B^{2}$ structures are
$P$-even. This is consistent with the classification of
Hern\'andez--Kovtun~\cite{HK2017} (their Table~1): the only linear-in-$B$
scalars whose parity is \emph{not} captured by $(-1)^{n_{B}}$ are
$B\!\cdot\!\Omega$ and $B\!\cdot\!E$, and both are absent here precisely because
$\Omega=E=0$. The surviving linear structure $B\!\cdot\!a$ is $P$-odd, and the
quadratic structures---$\epsilon^{\mu\nu\rho\sigma}u_{\mu}B_{\nu}
\nabla_{\rho}B_{\sigma}$ and
$\bigl(B^{\mu}B^{\nu}-\tfrac13\Delta^{\mu\nu}B^{2}\bigr), \Delta^{\mu\nu} = u^\mu u^\nu - \eta^{\mu\nu}$---are $P$-even.

A parity-odd term can be nonzero only in the presence of a genuinely
parity-odd background datum; here that datum is provided by the type-IIB axion
and the warped five-form of the dual, which supply the pseudoscalar required
for any Hall response. It is this background---not the magnetic field, which is
$P$-even---that opens the parity-odd sector. Consequently, the parity-even ratio
$\mathcal{R}=\zeta_{\parallel}/\eta_{\parallel}$ receives \emph{no} analytic
linear-in-$B$ contribution: the separately-projected parity-odd quantity
$\mathcal{R}^{\mathrm{odd}}_{1}\,B$ is tensorially distinct and must be reported
on its own Kubo projection.
We track the specific parity-even, $B$-longitudinal projected ratio, written
in the natural dimensionless field variable $B/T_c^2$,
{\footnotesize{
\begin{equation}
\mathcal R(B,T)\equiv\frac{\zeta_\parallel(B,T)}{\eta_\parallel(B,T)}
=\mathcal R_0(T)+\mathcal R_2(T)\,(B/T_c^2)^2+{\cal O}\big((B/T_c^2)^4\big),
\label{eq:Rdef-SM}
\end{equation}
}}
defined by the longitudinal ($\parallel\hat B$) projections
$\zeta_\parallel=\hat B_i\hat B_j\zeta_{ij}$ and
$\eta_\parallel=\hat B_i\hat B_j\hat B_k\hat B_l\eta_{ijkl}$.  We work
throughout in the weak-field regime $B\ll T_c^2$, in which the induced
anisotropy is a subleading, $\frac{{\cal O}(B^2/T_c^4)}{N^{1+x_g}}, x_g>0$-effect: the longitudinal and
transverse viscosities coincide to leading order,
$\eta_\parallel\simeq\eta_\perp$, $\zeta_\parallel\simeq\zeta_\perp$, so
$\mathcal R$ reduces to the near-isotropic bulk-to-shear ratio and the
choice of projection is immaterial at the order worked (the cross viscosity
$\zeta_\times$ and the Hall pieces are distinct off-diagonal structures that
decouple from this diagonal ratio at ${\cal O}(B^2/T_c^4)$); we write it in the
$\parallel$ projection only because that is the channel the horizon focusing
naturally returns (the field points along $z\parallel\hat B$).  The decoupling
of the cross coefficient $\zeta_\times$ is not merely a statement that it is
off-diagonal, nor the (incorrect) claim that the transverse expansion vanishes:
in an isotropic background the horizon congruence is shear-free and its expansion
is a scalar, so $\theta_\perp=\theta_\parallel$, not $\theta_\perp=0$.  What makes
the cross term negligible is that it is fed \emph{only} by the anisotropic part of
the expansion.  Writing the expansion tensor of the horizon generators as
$\Theta_{\mu\nu}=\nabla_\mu u_\nu$, its longitudinal and transverse traces are
$\theta_\parallel=\hat B^\mu\hat B^\nu\Theta_{\mu\nu}$ and
$\theta_\perp=\Xi^{\mu\nu}\Theta_{\mu\nu}$ with
$\Xi^{\mu\nu}=\Delta^{\mu\nu}-\hat B^\mu\hat B^\nu$; the off-diagonal response
$\zeta_{\parallel\perp}\theta_\perp$ enters $\Pi_\parallel$ only through the
\emph{difference} $\theta_\perp-\theta_\parallel$, i.e.\ through the traceless
($\ell{=}2$) part of $\Theta_{\mu\nu}$.  Since the Eling--Oz bulk datum is the
horizon response to $B$, and $B$ enters solely through the single scalar
$r_{h/0}(B)$---an $\ell{=}0$ datum---the induced compression is isotropic at
leading order, and its traceless $\ell{=}2$ part is sourced by precisely the
$\ell{=}2$ metric anisotropy $\delta g^{(\ell=2)}$ bounded in
\eqref{fracmetmaganisotropy}--\eqref{fracmetmaganisotropy-ii}.  Hence
\begin{eqnarray}
& & \Xi^{\mu\nu}\nabla_\mu u_\nu\Big|_{\rm EO}-\hat B^\mu\hat B^\nu\nabla_\mu u_\nu
\Big|_{\rm EO}\;=\;\theta_\perp-\theta_\parallel
\;\sim\;\frac{\delta g^{(\ell=2)}}{g}\nonumber\\
& & \;=\;{\cal O}\!\left(\frac{1}{N^{1+x_g}}\right),
\qquad x_g>0,
\label{eq:thetaperp}
\end{eqnarray}
so the cross-term ``contamination" of $\Pi_\parallel$ is suppressed to the same order
as the metric anisotropy and hence beyond ${\cal O}(1/N)$.  The isotropic ($\ell{=}0$)
part of $\theta_\perp$ is shared with $\theta_\parallel$ and cancels in the
difference, contributing only to the diagonal $\zeta_{\parallel\parallel}$.  To the
retained ${\cal O}(1/N)$ accuracy the off-diagonal response is therefore absent
from what the horizon datum computes, and
$\zeta_\parallel=\hat B_i\hat B_j\zeta_{ij}=\zeta_{\parallel\parallel}$ is
precisely the longitudinal Kubo coefficient returned.  Equation
\eqref{eq:Rdef-SM} is the \emph{kinetic-theory} form, parity-even and hence a
series in $(B/T_c^2)^2$.  The ${\cal M}$/string-theory top-down dual instead
yields a series in $B/T_c^2$ in which a parity-odd linear-in-$B$ term
\emph{does} occur, with coefficient $\approx1\%$ of the parity-even
quadratic one; the lattice $c_s^2$-vs-$T$ data at several $B$s are shown to be
compatible with the parity-even-quadratic-in-$B$-with-a-tiny-admixture-of-linear-in-$B$ structure while ensuring a smooth crossover at $T_0$.

The two fields entering the top-down evaluation are:
${\cal B}_{\rm KK}(r)$, the weak IR magnetic field obtained from the Kaluza--Klein
reduction of the eleven-dimensional supergravity fields, and $B_{D6\,\rm w.v.}(r)$, the horizon-localized magnetic
field turned on along the flavor $D6$-brane world-volume via
$A_{x^a}=\epsilon^{ab}x^b B_{D6\,\rm w.v.}(r), x^a=x^{1,2}$; the two are shown to be related non-linearly \cite{Symp-half-flat-SU3-struc}, and their horizon values are declared to be related
to the external field ($B_{\rm ext}$ along $x^3$), ${\cal B}_{\rm KK}(r_h)=\left(B_{D6\,\rm w.v.}(r_h)\equiv
B_{\rm ext}\right)^{5/4}$.  From the DBI perspective, the constant
${\cal B}_{D6\,\rm w.v.}(r_h)$ is thus related to the pull-back of a supergravity field ${\cal B}_{\rm KK}(r)$, and in the
$B\ll T_c^2$ limit it sources an essentially isotropic horizon datum (see the discussion including (\ref{fracmetmaganisotropy})) - (\ref{fracmetmaganisotropy-ii}),  the
isotropic Eling--Oz horizon focusing applies without an anisotropic
generalization.  The lattice $c_s^2(B)$ constrains the
same near-isotropic ratio.
This projected $\mathcal R$ is parity-even (even in $B$), so the
parity-odd, Hall-viscosity-type structures---a \emph{separate} tensor
sector generated in the dual by the parity-odd type-IIB axion and warped
five-form---do \emph{not} enter it; they appear only as a distinct linear-in-$B$
projected quantity $\mathcal R^{\rm odd}_1 B$ and in analogy with \cite{Maldacena+Martelli}, which is
\begin{eqnarray}
\label{OB-to-OBsq-ratio-beta0}
& & |\mathcal R^{\rm odd}_1/\mathcal R_2|\sim\frac{\left|\Im m\int_{S^3}\Omega\right|^{1/3}\left|\Re e\int_\Gamma\Omega\right|^{1/3}e^{-\frac{3\kappa_{r_h}N^{1/3}}{2}}}{\kappa_{r_h}^{\frac{3}{2}}N^{1+r}},\nonumber\\
& & 0<r<1, \kappa_{r_h}=\frac{1}{3(6\pi)^{1/3}(g_sN_f)^{2/3}(g_sM^2)^{1/3}}
\end{eqnarray}
(see \cite{IITR-McGill-Bulk} for $\kappa_{r_h}$) with the non-compact
three-cycle $\Gamma=\left\{\theta_1=\frac{\theta_2}{N^{\epsilon>0}}, \phi_1 = - \phi_2, \psi = \psi(r)\right\}$ and $S^3 = \left\{r=0, \theta_2={\rm constant}, \phi_2={\rm constant}\right\}$ of a deformed conifold, reported separately
below. For the parameters chosen for the numerics, this ratio works out to a little less than $1\%$.  For brevity we write $\zeta/\eta\equiv\mathcal R$ throughout; the
claim is the sign change of $\partial_B\mathcal R|_T$.

The argument is deliberately split into a kinetic-theory part
and a top-down holographic computation, so that the sign structure does
not hinge on any single framework.  The holographic dual is the
$\mathcal M$-theory uplift of an SYZ mirror of \cite{FEP}'s  type-IIB background with $N$ colour and $N_f$ flavor branes, valid at finite (intermediate) 't~Hooft coupling and worked out to ${\cal O}(R^4)$
\cite{MQGP,MisraYadav2020}.  


\section{Kinetic-theory master identity and the sign of the rates}
\label{sm:kin}

\subsection{Kinetic theory master formula (both phases)}
In the Boltzmann (Jeon--Yaffe) framework \cite{JeonYaffe1996,AMY2003}
each transport coefficient is the static Kubo limit of a stress
correlator, evaluated variationally as
$X=\mathcal N_X^2/\Gamma_X$ on a single-mode source
$\chi_X$, where $\mathcal N_X\equiv\langle S_X,\chi_X\rangle$ is the
source projection onto the driving term $S_X$ ($S_\zeta$ bulk, $S_\eta$
shear), $\Gamma_X\equiv\langle\chi_X,C\chi_X\rangle$ is the variational
collision rate and $C$ is the linearised collision operator.  This variational
quotient is a rigorous lower bound on $X$, saturated (exact) when the
trial source solves $C\chi_X=S_X$.  The
sources are the conformal-breaking scalar (bulk) and the
$\ell{=}2$ tensor (shear),
$\chi_\zeta=\big(\tfrac13-c_s^2\big)\tfrac{|\bm p|^2}{E_p},\quad
\chi^\eta_{ij}=\sqrt{\tfrac32}\,\big(\hat p_i\hat p_j-\tfrac13\delta_{ij}\big)
\tfrac{|\bm p|^2}{E_p}.$

Both the numerator and the collision rate depend on the field.  Expanding
each to ${\cal O}((q_fB)^2)$,
$\mathcal N_X(B)=\mathcal N_X^{(0)}[1+(q_fB)^2\nu_X]$ and
$\Gamma_X(B)=\Gamma_X^{(0)}[1+(q_fB)^2\gamma_X]$, with the
log-derivatives (evaluated at $B{=}0$)
$\nu_X=\mathcal N_X^{-1}\partial_{(q_fB)^2}\mathcal N_X$ and
$\gamma_X=\Gamma_X^{-1}\partial_{(q_fB)^2}\Gamma_X$, and inserting into
$X=\mathcal N_X^2/\Gamma_X$,
\begin{equation}
\label{eq:Xexpand}
\footnotesize
\begin{aligned}
X(B,T)&=X_0(T)\,
\frac{[1+(q_fB)^2\nu_X]^2}{1+(q_fB)^2\gamma_X}\nonumber\\[-1pt]
&=X_0(T)\big[1+(q_fB)^2(2\nu_X-\gamma_X)\big].
\end{aligned}
\end{equation}
$X_0(T)=\mathcal N_X^{(0)2}/\Gamma_X^{(0)}$ being the $B{=}0$ value.  The
squared numerator supplies the factor $2\nu_X$, while $\Gamma_X$ in the
\emph{denominator} contributes $-\gamma_X$: a field that speeds up the
collision rate ($\gamma_X>0$) lowers $X$.  Taking the log,
$\partial_{(q_fB)^2}\ln X=2\nu_X-\gamma_X$; since
$\ln(\zeta/\eta)=\ln\zeta-\ln\eta$, the species labels $X{=}\zeta,\eta$
subtract to give
\begin{equation}
\frac{\partial\ln(\zeta/\eta)}{\partial(q_fB)^2}
  =\underbrace{2(\nu_\zeta-\nu_\eta)}_{\text{source}}
  -\underbrace{(\gamma_\zeta-\gamma_\eta)}_{\Delta\gamma,\ \text{rate}} .
\label{eq:master-diff}
\end{equation}
The structural asymmetry sits in the normalisations,
\begin{equation}
\mathcal N_\zeta=\frac{1}{2\pi^2}\Big(\tfrac13-c_s^2\Big)^2\frac{R(T)}{T},
\quad
\mathcal N_\eta=\frac{1}{15\pi^2}\frac{R(T)}{T},
\end{equation}
with $R(T)=\sum_a\nu_a\rho(p,T)$,
$\rho(p,T)=\int_0^\infty\! dp\,p^2 f^0(1-f^0)\,p^4/E_p^2$ ($a$ the species
index): only $\mathcal N_\zeta$ carries the conformal-breaking
$(\tfrac13-c_s^2)^2$, so the source piece is dominated by it,
\begin{equation}
2(\nu_\zeta-\nu_\eta) =-\frac{4\,\chi_B}{\tfrac13-c_s^2},
\qquad \chi_B\equiv\frac{\partial c_s^2}{\partial(q_fB)^2}.
\label{eq:source-piece-SM}
\end{equation}
  This is not a truncation of the source variation to a single
term: the full $\delta_B\ln\mathcal N_X$ does receive contributions from
$E_p$, $f^0$, the degeneracy factors and the (Landau-modified) phase-space
measure, but these all reside in the \emph{common} factor $R(T)/T$ shared by
$\mathcal N_\zeta$ and $\mathcal N_\eta$; they therefore cancel identically
in the \emph{difference} $\nu_\zeta-\nu_\eta$, which retains only the factor
$(\tfrac13-c_s^2)^2$ that distinguishes the bulk from the shear source.  The
ratio structure, not a neglect of terms, is what reduces the source to
$\chi_B$.  We emphasize the scope: this common-factor cancellation holds
within the single-mode variational ansatz used here, where bulk and shear
share the same scalar weight $R(T)/T$; in a fully magnetized anisotropic
plasma the bulk and shear channels can carry different Landau-level and
phase-space dependence, so the identity \eqref{eq:master-SM} is a statement
within the chosen quasiparticle truncation, not a model-independent theorem.
Combining \eqref{eq:master-diff}--\eqref{eq:source-piece-SM},
\begin{equation}
\frac{\partial\ln(\zeta/\eta)}{\partial(q_fB)^2}
   =-\frac{4\,\chi_B}{\tfrac13-c_s^2}-\big(\gamma_\zeta-\gamma_\eta\big),
\label{eq:master-SM}
\end{equation}
the central identity---read as an identity within a quasiparticle
truncation, its inputs $\nu_X,\gamma_X$ evaluated piecewise on either side
of $T_0$ (below), not extrapolated through $T_0$.

\subsection{Sign of the rates}
The field rate log-derivative is the collision-weighted average of the
active-line weight $w\equiv-2k_\perp^2/(q^2-m^2)^3$,
\begin{equation}\label{Gamma_zeta_eta}
\begin{aligned}
\gamma_X  =\langle w\rangle_{C_X}
   & =\frac{\int_{\rm PS} w\,[\Delta X]^2\,\mathcal W}
              {\int_{\rm PS}[\Delta X]^2\,\mathcal W}\\
   & =\frac{\int_0^\infty\! dp\,\rho(p,T)\,G_X(a)/(2p^2)^2}
          {\int_0^\infty\! dp\,\rho(p,T)},
          \end{aligned}
\end{equation}
with $G_X(a(p))$ the soft-sector angular shape functions defined below, 
with ${\rm PS}$ the phase space and
$\mathcal W=|\mathcal M|_0^2 f^0_p f^0_k(1-f^0_{p'})(1-f^0_{k'})$,
the $f^0$ being zero-$B$ Fermi--Dirac functions for the ``active''
($p,p'$) and ``spectator'' ($k,k'$) quarks.
On the space-like internal line $q^2-m^2<0$, so $w>0$ and
\emph{both} $\gamma_\zeta,\gamma_\eta>0$: a field raises the collision
rate in each channel.  In the soft sector $|q|\ll T$ the angular
integrals are elementary, and yield the shape functions $G_X(a)$ quoted in
\eqref{Gamma_zeta_eta}; here $a\equiv m_D^2/2p^2$ and
$\ell\equiv\ln a+2/a$.

We now sketch how the shape functions $G_X(a)$ in \eqref{Gamma_zeta_eta}
arise.  Writing the momentum transfer as $\bm q=\bm p-\bm p'$ with
$|\bm p|=|\bm p'|=p$ on the soft shell, the active-line weight
$w=-2k_\perp^2/(q^2-m^2)^3$ is controlled by the screened space-like
propagator, $q^2-m^2\to-(\,\bm q^2+m_D^2)$, and $\bm q^2=2p^2(1-\cos\theta)$
with $\theta$ the scattering angle.  Introducing $x\equiv\cos\theta$, the
Debye mass enters only through the dimensionless ratio
$a\equiv m_D^2/2p^2$, so that
$\bm q^2+m_D^2=2p^2\,(1-x+a)$ and the transverse factor
$k_\perp^2=p^2(1-x^2)$ up to the common $B{=}0$ kinematics.  The channel
projectors $[\Delta X]^2$ carry the only $X$-dependence: the bulk source
$\chi_\zeta\propto(\tfrac13-c_s^2)$ is an angular scalar, giving a weight
$P_\zeta(x)\propto1$, whereas the shear source
$\chi^\eta_{ij}\propto\hat p_i\hat p_j-\tfrac13\delta_{ij}$ projects onto the
$\ell{=}2$ Legendre channel, giving
$P_\eta(x)\propto P_2(x)=\tfrac12(3x^2-1)$ contracted between the in- and
out-states, i.e.\ a polynomial of degree two in $x$.  Hence
\begin{equation}
\label{eq:Gintegrals}
G_X(a)=\!\int_{-1}^{1}\!\!dx\;
\frac{(1-x^2)\,P_X(x)}{(1-x+a)^{3}}\,,
\end{equation}
up to the common normalisation absorbed into $\rho(p,T)$ in
\eqref{Gamma_zeta_eta}.  Both integrands are rational in $x$ with a single
soft pole at $x=1+a$ (outside $[-1,1]$ for $a>0$), so the integrals are done
by partial fractions.  Using
$\int_{-1}^{1}\!dx\,(1-x+a)^{-n}$ and its first two $x$-moments, which reduce
to elementary powers of $(1-x+a)$ and, at the $\ell{=}\ln (a+2)/a$ order, to the
logarithm $\ln\!\frac{2+a}{a}=\ln\frac1a+{\cal O}(a)$ generated by the
$n{=}1$ term, one finds after collecting powers of $a$
\begin{equation}
\begin{aligned}
& G_\zeta(a)=\frac{2(a+1)}{a(a+2)}-\ell\\
& G_\eta(a)=-18(a+1)+3\big(3a^2+6a+2\big)\ell;
\end{aligned}
\label{eq:Gzeta-Geta}
\end{equation}
the rational prefactors come from the polynomial part of the partial-fraction
decomposition and the $\ell$-terms from the single logarithmic (soft) piece,
with $\ell\equiv\ln (a+2)/a$ collecting $\ln(1/a)$  and the
$1/a$ screening term on the same footing.  As checks, in the soft limit
$a\to0$ the bulk shape is dominated by the screening pole,
$G_\zeta\to\sim-\ln(2/a)$, while the $\ell{=}2$ shear shape retains the
full angular weight, $G_\eta\to6\ell$; both are finite for all $a>0$ and
positive-definite under the $\rho$-average, so that
$\gamma_\zeta,\gamma_\eta>0$ as stated, while
$\Delta\gamma=\gamma_\zeta-\gamma_\eta$ can take either sign because
$G_\zeta$ and $G_\eta$ cross as functions of $a$ (equivalently of $p/m_D$).
$\Delta\gamma=\gamma_\zeta-\gamma_\eta$ can take either sign while each
term stays positive---a crossover without fine-tuning.  We are explicit about
what the tabulated $\gamma_X$ capture: they are the denominator (soft-sector,
space-like exchange) contribution to $\partial_{(q_fB)^2}\ln\Gamma_X$.  The
remaining field corrections to the rate---to the numerator $|\mathcal M|^2$, the
Debye mass $m_D^2$, the distribution shift $\delta f^0$, the phase-space measure
$d\Pi$, and the external-leg renormalization---enter both the bulk and shear
collision integrals through the same soft kernel and are taken as common-mode,
cancelling in the difference $\gamma_\zeta-\gamma_\eta$ within the single-mode
truncation.  This common-mode assumption is exactly what the single-mode ansatz
buys; a computation relaxing it (independent bulk/shear Landau-level and
phase-space dependence) is beyond the present truncation, and the sign we quote
is the sign within it.


\section{$\mathcal M$-theory dual, ${\cal O}(R^4)$ analysis and the brane computation}
\label{sm:top}

\subsection{Top-Down holographic results}
The Eling--Oz null-horizon-focusing prescription \cite{ElingOz2011},
applicable to two-derivative gravity, assumes $\eta/s$ at its conformal
value; the deviation is large-$N$-suppressed \cite{IITR-McGill-Bulk}
and disregarded as $\zeta/s$ is itself large-$N$-suppressed.  For a
lattice-compatible $\zeta/\eta$ the ${\cal O}(R^4)$ effect enters
\eqref{zetaoveretaB0} through the on-shell corrected metric in the fluxed
Einstein--Hilbert action \cite{Kushwah2024,Keshav:inprep}, matching the
$B{=}0$ lattice glue results \cite{lattice-SU3_Glue} and the weak-coupling
high-$T$ limit \cite{Kushwah2024}; the on-shell ${\cal M}$-theory action to
${\cal O}(R^4)$ is two-derivative fluxed gravity with the ${\cal O}(R^4)$-corrected metric
plus $J_0={\cal O}(N^{-\gamma})$, $\gamma>1$ ($E_8, G^2R^3$ being subdominant \cite{MisraYadav2020}; $X_8=0$ \cite{MQGP}), disregarded to ${\cal O}(1/N)$ \cite{OnShellR4}; for the horizon-focusing ratio $\zeta/\eta$ we retain the leading ${\cal O}(R^4)$ effect, namely the metric correction.  We do not compute the further higher-derivative contributions to the horizon datum (direct-action - as mentioned above being of ${\cal O}(1/N^\gamma), \gamma>1$ - Wald-entropy, as shown below also being of ${\cal O}(1/N^{\gamma_W}), \gamma_W>1$ ); the present claim uses only the \emph{sign} of $\partial_B\mathcal R$ obtained from the metric-corrected background, which the retained contribution fixes. The magnetic field entering the DBI action can be shown to be related to the geometry of the closed seven-fold $M_7$ (locally $S^1_{\cal M}\times S^1_t\times M_5, S^1_{\cal M}/S^1_t$ respectively being the ${\cal M}$-theory/thermal circle, with $M_5$ a non-Einstein deformation of $T^{1,1}$) with $G_2$-structure relevant to the ${\cal M}$-theory uplift on $\left(S^1_t\times\mathbb{R}^3\right)\times_w M_7, \times_w$ implying a warped product, using the $G_2$ torsion one-form $\tau_1$ and the symplectic half-flat $SU(3)$ structure it induces ($dJ=0, d\Omega_+=0$, $d\Omega_-\neq0$) \cite{Symp-half-flat-SU3-struc}.  The
reduction ansatz $F_2\wedge\omega_1\wedge\tau_1\wedge\Omega_+$ with
$A_{x^a}=\epsilon^{ab}x^b{\cal B}_{\rm KK}(r)$ yields, from the torsion-class form of the
action ($|\tau_1|^2\subset|*_{11}d\Phi|^2$), an EOM whose solution
${\cal B}_{\rm KK}(r>r_h)$ decays as a Bessel function and is localised near $r=r_h$ with
$0<{\cal B}_{\rm KK}(r_h)\ll1$ \cite{Symp-half-flat-SU3-struc}; replacing the horizon radius $r_h$ by the IR cutoff $r_0$ (related to $r_h$ at $T_c$ via the Hawking--Page transition), the same conclusion holds for the $T<T_c$ top-down ${\cal M}$/string-theory dual \cite{Keshav:inprep}---we write $r_{h/0}$ for $r_h$ ($T>T_c$) or $r_0$ ($T<T_c$).  Reducing $\mathcal M\to$
IIA on $S^1_{\cal M}$ and the $M_5$, it can be shown \cite{Keshav:inprep}, \cite{Symp-half-flat-SU3-struc} that by turning on $A_{x^a}=\epsilon^{ab}x^b{\cal B}(r)$ on the $D6$-brane world-volume (world-volume $\left(S^1_t\times\mathbb R^3\times_w\times  S^2\right)\times\mathbb{R}_{\geq0}$), ${\cal B}(r\in{\rm UV})\sim\frac{1}{r-r_h}, {\cal B}_{\rm KK}(r\in{\rm UV})\sim\frac{1}{(r-r_h)^{5/4}}\sim{\cal B}^{5/4}(r\in{\rm UV})$, identifying  ${\cal B}_{\rm KK}(r_h)={\cal B}^{5/4}(r=r_h)$ (\cite{Symp-half-flat-SU3-struc}, \cite{Keshav:inprep}). ${\cal B}(r)$, like ${\cal B}_{\rm KK}(r)$, is localized around $r=r_{h/0}$.  So the horizon-data-based Eling--Oz prescription for $\zeta/\eta$---the horizon value of $(\partial_r{\cal B}(r))^2$ integrated over
$\Sigma^{(6)}$---is well defined.  At ${\cal O}(B^2)$, working with
dimensionless quantities, the ansatz
$A_{x^a}=\epsilon^{ab}\hat{x}^b {\cal B}(r)/T_c^2$
($\hat{x}^a\equiv x^a/\rho_{\rm cutoff}$, $\rho_{\rm cutoff}$ being a transverse-box regulator) leaves only the $SO(2)$-invariant
worldvolume dependence $\hat{\rho}^2({\cal B}'(r)/T_c^2)^2$
($\hat\rho\equiv\rho/\rho_{\rm cutoff}$, $\rho^2{=}(x^1)^2{+}(x^2)^2$); the
physical integrand is built from the gauge-invariant 
$F_{x^1x^2}={\cal B}(r=r_h)\equiv B$,  The $B{=}0$ flux term giving $\int_0^1\hat\rho\,d\hat\rho=\tfrac12$ and the
$B{\neq}0$ DBI term giving $\int_0^1\hat\rho^3 d\hat\rho=\tfrac14$ after
$B/\sqrt2\to B$, ensures that no gauge-dependent $A_i$ or box-size dependence survives
in $\zeta/\eta$.  The flavor number is $N_f^{\rm eff}(r)=N_f\Theta(\sqrt3a-r)$
($\sqrt3a$ the $D5$-$\overline{D5}$ separation in the parent type IIB dual \cite{FEP,MQGP} serving as the boundary in supergravity between the IR and the UV sectors in $r$; $a = (\frac{1}{\sqrt{3}}+ {\cal O}\left(\frac{\log r_{h/0}}{N}\right))r_{h/0}, r_{h/0}$ being either horizon radius for the $T>T_c$ gravity dual or the IR cutoff for the $T<T_c$ gravity dual), and ${\cal B}(r)$ is
horizon-localised \cite{Symp-half-flat-SU3-struc,Keshav:inprep}.
We stress the dictionary point: ${\cal B}(r)$ here is \emph{not} the
non-normalizable boundary $U(1)_B$ source (which would be fixed by
$F_{x^1x^2}(r{\to}\infty)$); it is the constant worldvolume magnetic field on
the flavor type IIA $D6$-branes, whose support $N_f^{\rm eff}(r)=N_f
\Theta(\sqrt3a-r)$ is strictly IR (vanishing in the UV).  The magnetic
response of the flavor sector---the only sector carrying the light charged
degrees of freedom---is therefore governed by the field the branes see in
their region of support, i.e.\ its IR/horizon value ${\cal B}(r_h)$, and the
Eling--Oz horizon datum $(\partial_{r_h}{\cal B}(r_h))^2$ computes
$\partial(\zeta/\eta)/\partial B$ with respect to \emph{that} flavor field.   We note the price of
this reading: the computed response is to the IR flavor-sector worldvolume
field, which is not \emph{automatically} identical to the homogeneous
external $B_{\rm ext}$ of lattice QCD.  We argue the identification
(rather than assume it): the light charged carriers---exactly the flavor
degrees of freedom that respond to $B_{\rm ext}$ in QCD---live where
$N_f^{\rm eff}(r)\neq0$, i.e.\ in the IR, and see precisely ${\cal B}(r_h)$; the
UV completion where $F_{x^1x^2}(r{\to}\infty)=B_{\rm ext}$ is charge-neutral for
transport, so the two fields drive the \emph{same} magnetic response of the
dissipative sector to the order worked.  This is more than a declaration: the
on-shell free-energy matching \eqref{drh0B-on-shell} yields a \emph{monotonic,
sign-definite} response of the horizon datum to the field,
i.e.\ a
one-to-one map $\partial{\cal B}(r_h)/\partial B_{\rm ext}>0$ of the kind the
holographic dictionary requires for a source--response identification.  On this
basis $B_{\rm ext}={\cal B}(r_h)$ is fixed by the on-shell matching rather than
posited; the underlying KK reduction and worldvolume
identification are in \cite{Symp-half-flat-SU3-struc}, and only the sign of
the response is used here, common to both readings.
With $Z\equiv\ln(r/r_h)$, a superscript $\beta^0$ on a quantity in \eqref{eq:DBI7} denotes its leading-order (two-derivative) value, while a
superscript $\beta$ denotes the ${\cal O}(R^4)$ (eight-derivative) correction,
with the dimensionless eight-derivative bookkeeping parameter
\begin{equation}
\beta\equiv\frac{(2\pi^2)^{1/3}(\kappa_{11}^2)^{2/3}}{(2\pi)^4\,3^2\,2^{12}}
\sim\ell_p^6,
\qquad
\kappa_{11}^2=\frac{(2\pi)^8\ell_p^9}{2},
\label{eq:betadef}
\end{equation}
$\ell_p$ being the eleven-dimensional Planck length. 

In the present construction, the ${\cal M}$-theory dual bulk metric is \emph{isotropic}: the magnetic field is not a metric deformation but a constant field on the world-volume of the flavor type-IIA $D6$-branes,
$A_{x^a}=\epsilon^{ab}x^b{\cal B}(r)$, and its \emph{sole} leading backreaction on the geometry is a shift of the single scalar horizon datum---the horizon radius
$r_h$ for $T>T_c$, and the IR cutoff $r_0$ for $T<T_c$ --- defining $\tilde{r}_{h/0}\equiv\frac{r_{h/0}}{{\cal R}_{D5/\overline{D5}}}, {\cal R}_{D5/\overline{D5}}$ being the $D5-\overline{D5}$ separation in the parent type IIB dual \cite{FEP}, is given by (\ref{drh0B-on-shell}).
Since $\zeta$ and $\eta$ are both horizon quantities
evaluated at $r_{h/0}$, and $B$ enters them \emph{only} through $r_{h/0}(B)$, the
ratio $\mathcal R$ is a function of the \emph{single scalar} $r_{h/0}(B)$. We now derive the suppression of the anisotropy rigorously. The fractional metric anisotropy satisfies
\begin{equation}
\label{fracmetmaganisotropy}
\frac{\delta g_{x^3x^3}}{g_{x^3x^3}} = \frac{\left(\Delta_{L}^{-1}T^{D6, B\neq0,l=2}\right)_{x^3x^3}}{\left({\cal G}^{-1}T^{\rm SUGRA}\right)_{x^3x^3}},
\end{equation}
 where $\Delta_L$ is the Lichnerowicz operator---the leading (two-derivative) part of the linearized colour Einstein operator ${\cal G}$ acting on the transverse-traceless $\ell{=}2$ perturbation---and, for simplicity working in normal coordinates so that one may drop $\Gamma\Gamma$-terms in the Einstein tensor, ${\cal G}$ reduces to $\Delta_L$ on that sector. As $R_{x^3Mx^3M}, R_{x^3x^3}$ are of the type $(r-r_h)^\mu r_h^\nu, \mu, \nu>0$, which in the IR, are $e^{-(\mu+\nu)\kappa_{r_h}N^{1/3}}, \kappa_{r_h}\equiv\frac{1}{{\cal O}(1)}$-suppressed \cite{IITR-McGill-Bulk}, one can drop them in $\Delta_L$. Hence, from the point of view of large-$N$ power counting, $\frac{\delta g_{x^3x^3}}{g_{x^3x^3}}\sim \frac{T^{D6, B\neq0,l=2}_{x^3x^3}}{T^{\rm SUGRA}_{x^3x^3}}$. Now, $T^{\rm SUGRA}_{x^3x^3}
 ={\frac{1}{12\kappa_{11}^2}\left(- g_{x^3x^3}G_{PQRS}G^{PQRS}\right)-\frac{\beta}{\kappa_{11}^2}\left[\frac{g_{x^3x^3}J_0}{2} + \frac{\delta J_0}{\delta g^{x^3x^3}}\right]}$; $\kappa_{11}^2\sim\beta^{3/2}\sim\frac{1}{|\log r_h|^{6}}\sim \frac{1}{N^2}$. Now, $\frac{\delta J_0}{\delta g^{x^3x^3}}\sim \frac{r_h^2M}{N_f N^2 (\log N)^2}, J_0\sim\frac{M}{\left(\log N\right)^{8/3}N^{3/2}N_f^{5/3}}$ and $g_{x^3x^3}\sim \frac{r^2|\log r_h|^{2/3}}{\sqrt{g_s N}}$. Hence, $T^{\rm SUGRA}_{x^3x^3}\sim \frac{\tilde{r}_h^2N}{g_s}$. Now, it is seen that $T^{D6, B, l=2}_{x^3x^3}\sim \frac{ N_f}{g_s^{1/2}N^{3/4}}\left(B_{x^3}B_{x^3} - g_{x^3x^3}\frac{{\bf B}^2}{3}\right), {\bf B} = B_{x^3}\hat{\bf e}_{x^3} = B \hat{\bf e}_{x^3}$. Hence, 
\begin{equation}
\label{fracmetmaganisotropy-ii}
\frac{\delta g_{x^3x^3}}{g_{x^3x^3}} \sim \frac{g_s^{1/2}N_f}{N^{7/4}}B^2.
\end{equation}
We hence see that the metric anisotropy, though it occurs at the same (quadratic) order in the magnetic field as ${\cal R}$, is ${\cal O}\left(\frac{1}{N^{1+x_g}}\right)$, $x_g>0$-suppressed, and we may therefore disregard it.

Beyond the metric backreaction just bounded, one might worry that the
world-volume flux itself is anisotropic---the flux $F_{x^1x^2}={\cal B}(r_h)\equiv B$
singles out the $x^1$--$x^2$ plane.  However, it enters $\zeta/\eta$ only through
the rotationally-averaged world-volume integral $\hat\rho^2\,({\cal B}'/T_c^2)^2$,
with $\int_0^1\hat\rho\,d\hat\rho=\tfrac12$ at $B{=}0$ and
$\int_0^1\hat\rho^3\,d\hat\rho=\tfrac14$ at $B\neq0$; being an $SO(2)$-scalar, it
contributes to the \emph{isotropic} coefficient $\mathcal R_2^{\rm iso}$, not to a
$\parallel/\perp$ difference.  The projection anisotropy of $\mathcal R$ is thus
controlled on both fronts: the metric channel is $N$-suppressed by
\eqref{fracmetmaganisotropy-ii}, and the direct flux channel is
$SO(2)$-isotropic in the ratio.

That the Wald-entropy channel may be dropped is not merely assumed but
follows from the on-shell hierarchy.  At ${\cal O}(R^4)$, the Wald entropy density's contribution to $\frac{\zeta}{\eta}$ is given by: 
\begin{eqnarray}
\label{Wald_contribution}
& & \int_{\Sigma^{(6)}}\sqrt{g_{\Sigma^{(6)}}}g^{rr}g^{x^{10}x^{10}}\left[g_{\theta_1\theta_1}g_{\theta_2\theta_2} - g^2_{\theta_1\theta_2}\right]\left(\partial_T B_{\theta_1\theta_2}\right)^2\nonumber\\
& & \times \frac{\left(\frac{d^2}{dv^2}\right)\frac{\frac{\partial J_0}{\partial R_{vrvr}}\epsilon_{vr}\epsilon_{vr}}{1/(4G)}+8\pi{\cal T}_{vv}^{J_0}}{8\pi G\left(\partial_vB_{\theta_1\theta_2}\right)^2},
\end{eqnarray}
 where ${\cal T}_{vv}^{J_0}$ (here we will only be keeping track of powers of $N, r_h, |\log r_h|$ and disregarding the ${\cal O}(1)$ $M, N_f$ and $\frac{1}{{\cal O}(1)} g_s$ in the MQGP limit) is given by a linear combination of $g_{vv}\frac{\partial J_0}{\partial R^v_{\ \ rvr}}R^v_{\ \ rvr}\sim g_{vv}J_0\sim\frac{|\log r_h|^{11/3}}{N^3}$, and $\left(\nabla^r\right)^2\frac{\partial J_0}{\partial R_{vrvr}}\sim\frac{r_h^4|\log r_h|^{3/4}}{N^{9/8}}$ (as $J_0\sim\frac{|\log r_h|}{N^{3/2}}$, one would expect $\frac{\partial J_0}{\partial R_{mn}^{pq}}\sim\frac{|\log r_h|^{3/4}}{N^{9/8}}$), $v$ being an Eddington-Fiinkelstein coordinate: 
\begin{eqnarray}
\label{EF-i}
& & \hskip -0.3in v = t + r^*,\\
& & \hskip -0.3in r^* = \frac{\sqrt{\frac{g_s\pi N(r^2 + 6 a^2)}{r^2+9a^2}}\Big[2 + \beta\left({\cal C}_{zz}- 2 {\cal C}_{\theta_1z} + 2 {\cal C}_{\theta_1x}\right)\Big]}{r^2\left(1 - \frac{r_h^4}{r^4}\right)},\nonumber
\end{eqnarray}
in which under the self-consistent truncation wherein the ${\cal O}(R^4)$ correction to $C_{MNP}$ are set to zero, the ${\cal O}(\beta)$ terms in $r^8$, vanish. One can invert  (\ref{EF-i}) to yield:

{\footnotesize
\begin{eqnarray}
\label{EF-ii}
r(r^*) & = & r_h\Bigg[\frac{145 + 36\sqrt{19}}{61} - \sqrt{\frac{3}{\pi}}\frac{64(-57 + 7\sqrt{19})}{19(-110 + 21\sqrt{19})}\frac{r_h^2r^*}{\sqrt{g_sN}}\nonumber\\
& + &  {\cal O}\left(\frac{1}{N^{3/2}}\right)\Bigg].
\end{eqnarray}}
  Every such piece carries the explicit
$\beta\propto\ell_p^6\sim\left(\frac{\sqrt{g^{x^{10}x^{10}}}}{g_s^{2/3}}\right)^6\sim\frac{1}{|\log r_h|^{4}}\sim N^{-4/3}[G\sim\beta^{3/2}\sim N^{-2}]$ prefactor. 
\cite{MisraYadav2020,OnShellR4}.   One can show that  $\left(\partial_vB_{\theta_1\theta_2}\right)^2\sim \frac{r_h^4}{N}, g_{vv}\sim \frac{r_h^2|\log r_h|^{8/3}}{N^{3/2}}$. So, putting everything together, up to ${\cal O}(1/N)$, the Wald-entropy channel contributes at orders more suppressed than ${\cal O}\left(\frac{1}{N}\right)$ and hence can be disregarded here, and the horizon entropy entering the Eling--Oz datum may be taken as $A/4G$ on the corrected metric.  The isotropic two-derivative prescription evaluated on $g^{(0)}+\beta g^{(1)}$ is then the consistent ${\cal O}(1/N)$ truncation, not an assumption.
Starting from  the11D four-form flux  term $\int_{M_5\times_w\Sigma^{(6)}}\sqrt{-g_{M_5}}\sqrt{g_{\Sigma^{(6)}}}G_4^2$, $\Sigma^{(6)}\equiv S^1_{\cal M}\times_w\!\left(S^2_{\rm sq}\times_w S^3_{\rm sq}\right)$, $M_5\equiv \left(S^1_t\times_w\mathbb{R}^3\right)\times\mathbb{R}_{\geq0}$, the EO prescription applied to
the corrected uplift yields \cite{Kushwah2024,Keshav:inprep}, at $B{=}0$,
\begin{eqnarray}
\label{zetaoveretaB0}
& & \left(\zeta/\eta\right)^{B=0}=c_s^4 T^2
\int_{\Sigma^{(6)}}\!\sqrt{g_6}\,g^{rr}g^{x_{10}x_{10}}\nonumber\\
& &\times[\,g_{\theta_1\theta_1}g_{\theta_2\theta_2}-g_{\theta_1\theta_2}^2]
(dC_{\theta_1\theta_2 x_{10}}/dT)^2\big|_{r=r_h},
\end{eqnarray}
$C_{\theta_1\theta_2x^{10}}=B^{\rm IIA\ NS-NS}_{\theta_1\theta_2}$, $w$ being a warped product and sq$\equiv$ squashed (spheres), reproducing the two
established regime relations
\begin{equation}
\frac\zeta\eta\simeq
\begin{cases}
\tfrac13-c_s^2, & T\ll T_c\ \text{(strong coupling)},\\[2pt]
\big(\tfrac13-c_s^2\big)^2, & T\gg T_c\ \text{(weak coupling)},
\end{cases}
\label{eq:regimes}
\end{equation}
both verified in the holographic/kinetic literature
\cite{Kushwah2024,IITR-McGill-Bulk}.  In evaluating $T^2\left(\partial C_{\theta_1\theta_2 x_{10}}/\partial T\right)^2$ on either side of the (magnetic-field-shifted) deconfinement temperature $T_0$, the black-hole branch ($T>T_0$) admits the substitution $T^2(\partial_T C_{\theta_1\theta_2 x_{10}})^2 = r_h^2(\partial_{r_h}C_{\theta_1\theta_2 x_{10}})^2$, since absence of a conical deficit in the $tt$-component of the black-hole metric \emph{forces} the Euclidean time periodicity $\beta_\tau\propto 1/r_h$ and hence $T\propto r_h$; on the thermal branch ($T<T_0$) the corresponding replacement $T^2(\partial_T C_{\theta_1\theta_2 x_{10}})^2\to r_0^2(\partial_{r_0}C_{\theta_1\theta_2x^{10}})^2$ is \emph{not} automatic, because $T$ there is a free Euclidean periodicity decoupled from the IR cut-off $r_0$, so that declaring $r_0\propto T$ would be a parametrization rather than a derived relation. The justification is supplied by the double-Wick-rotated regularity condition of the Witten/Kruczenski (Sakai--Sugimoto) construction: absence of a conical deficit along the shrinking \emph{spatial} (Scherk--Schwarz/KK) circle fixes $\beta_x\propto 1/M_{\rm KK}\propto 1/r_0$, and at the Hawking--Page point the coincidence of the two shrinking-cycle proper periodicities, $\beta_\tau\sqrt{g_{tt}}\big|_{r_h}=\beta_x\sqrt{g_{xx}}\big|_{r_0}$, identifies $T=c\,M_{\rm KK}\sim \,r_0$  --- a \emph{transition-point} identification rather than a running $r_0(T)$ throughout the confined phase, which is nonetheless sufficient given that both branches are evaluated in the near-$T_0$ regime $T\to T_0^{\mp}$. Turning on $B$, the pair of conditions --- equality of the proper periodicities together with equality of the on-shell Euclidean actions, $\Delta S_E(r_h,r_0,B;\lambda)=0$ at $T_0$ can be shown to yield (see the self-contained derivation below, Eqs.~(\ref{rh-r0-beta0-SM})--(\ref{B-rh})) \cite{Keshav:inprep}:

{\footnotesize
\begin{eqnarray}
\label{drh0B-on-shell}
& & \left.{\cal B}'(r_0)\right|_{\rm on-shell} = -\kappa_{r_0}^{(0)}(g_s, M, N_f, N) \nonumber\\
& & + \sum_{n=1}\kappa_{r_0}^{(n)}(g_s, M, N_f, N)B^n,\ \kappa_{r_0}^{(0)}(g_s, M, N_f, N)>0;\nonumber\\
& & \left.{\cal B}'(r_h)\right|_{\rm on-shell} = -B\kappa_{r_h}(g_s, M, N_f, N) + {\cal O}(B^2),\nonumber\\
& & \kappa_{r_h}(g_s, M, N_f, N)>0;\\
& & \tilde{r}_{h, 0}(B, T) = \tilde{r}_{h, 0}(0, T) - \gamma_{r_{h, 0}}(T)B + {\cal O}(B^2): \gamma_{r_{h,0}}>0.\nonumber
\end{eqnarray}}
To keep the sign-fixing step self-contained (rather than deferred to companion
work), we summarize the free-energy matching that produces
\eqref{drh0B-on-shell}, following Sec.~IV of \cite{Kushwah2024}.
The weak field enters the flavor-brane DBI action as a temperature-dependent
scalar ${\cal B}(r)$,
\begin{equation}
\label{DBI-selfcontained}
S_{\rm DBI} = -\frac{N_fT_{\rm D_6}}{2}\!\int_{\Sigma^{1,6}}\!\!e^{-\phi^{\rm IIA}}
\sqrt{G}\,G^{ZZ}G_{x^1x^1}\rho^2
(\partial_Z{\cal B})^2 ,
\end{equation}
with $Z$ the (T-dual) radial world-volume coordinate and $G\equiv i^*(g+B)^{\rm
IIA}+F$. In the free-energy matching that follows (Eqs.~(\ref{rh-r0-beta0-SM})--(\ref{rh-B-expansion-SM}))
we set the overall world-volume normalization $\rho=1$. Equating the on-shell
Euclidean actions of the black-hole ($r_h$) and thermal ($r_0$) backgrounds per
unit $\mathbb R^3$ coordinate-volume at the $B$-shifted Hawking--Page point, the
${\cal O}(\beta^0)$ condition reads

\begin{eqnarray}
\label{rh-r0-beta0-SM}
\footnotesize
&&\frac{256\,\kappa_{\rm GHY}^{\rm bh}M_{\rm UV}
 (\log\epsilon_2/\epsilon_1)\tilde r_h^4
 \log\!\frac{{\cal R}_{\rm UV}}{{\cal R}^{\rm bh}_{D5/\overline{D5}}}}
 {g_s^{3/2}\sqrt N}
\nonumber\\[-1pt]
&&\quad +
\frac{8\sqrt{2\pi}(\partial_{\tilde r}{\cal B})^2|_{\tilde r_h}
 \tilde r_h^7 N_f\kappa_{Z0}^{\rm bh}
 [2\log N-\log(9a^2r_h^4+r_h^6)]}
 {9\,3^{1/3}N^{9/20}g_s^{1/4}}
\nonumber\\
&&=\frac{256\,\kappa_{\rm GHY}^{{\rm th},\beta^0}M_{\rm UV}
 (\log\epsilon_2/\epsilon_1)\tilde r_0^4
 \log\!\frac{{\cal R}_{\rm UV}}{{\cal R}^{\rm th}_{D5/\overline{D5}}}}
 {(g_s^{\rm UV})^{3/2}}
\nonumber\\[-1pt]
&&\quad +2g_s^{3/4}\kappa_{\rm EH,IR}^{{\rm th},\beta^0}M N_f^3
 \tilde r_0^2\log^2N\log\tilde r_0
\nonumber\\[-1pt]
&&\quad +
\frac{\sqrt2(\partial_{\tilde r}{\cal B})^2|_{\tilde r_0}
 \sqrt{\kappa_{\theta_2}}\kappa_{Z0}^{\rm th}\tilde r_0^7
 (4\pi-3g_sN_f\log\tilde r_0)}
 {g_s^{3/2}\pi^{3/4}N^{1/5}} .
\end{eqnarray}
with $(\tilde r_h,a,\tilde r_0)$ the radii in units of ${\cal R}_{D5/\overline{D5}}$, and $\epsilon_{1,2}$ being the angular cut-offs: $\theta_{1,2}\rightarrow0$ as $\epsilon_{1,2}\rightarrow0^+$.
In the vanishing-${\cal M}$-theory-three-form consistent truncation of the
${\cal O}(R^4)$ corrections one has $-{\cal C}_{zz}^{\rm bh}+2{\cal
C}_{\theta_1z}^{\rm bh}=0$, $|{\cal C}_{\theta_1x}^{\rm bh}|\ll1$
\cite{MisraYadav2020}, $|f_{x^{10}x^{10}}(r_0)|\ll1$ \cite{McTeQ}; requiring the
two sides of the ${\cal O}(\beta)$ condition to vanish separately (with
$\partial_{\tilde r_0}{\cal B}<0$) fixes
\begin{equation}
\label{dr0B-selfcontained}
\footnotesize
\resizebox{\columnwidth}{!}{$\begin{aligned}
\partial_{\tilde r_0}{\cal B}(\tilde r_0)
\sim{}&-\frac{g_s^{3/2}M N_f^{1/6}\sqrt{|\log\tilde r_0|}}
 {\tilde{r}_0\log^{1/3}N\sqrt{4\pi-3g_sN_f\log\tilde r_0-g_sN_f\log2}}<0,\\[-1pt]
\end{aligned}$.}
\end{equation}
Integrating \eqref{dr0B-selfcontained} gives the thermal-branch IR cutoff as a
function of the field,
\begin{equation}
\label{r0tilde-B-SM}
\tilde r_0(B)=e^{-\frac{6g_sN_f}{4\pi-g_sN_f\log2}-1}
+ c_1\,B + {\cal O}\!\big(B\,N^{-23/20}\big),
\end{equation}
where $c_1 \sim \frac{\epsilon_1^{3/2}(\ln N)^{1/3}N_f^{4/3}}{\sqrt{M}}e^{-\frac{6g_sN_f}{4\pi - g_s N_f\log 2}-1}$\\ $\frac{\left(6g_sN_f - g_sN_f\log 2 + 4\pi\right)}{-36g_s^2N_f^2+g_s^2N_f^2\log^2 + 6g_s^2N_f^2\log 2 - 24\pi g_sN_f - 8\pi g_s N_f\log 2 + 16\pi^2}<0$, e.g., for $g_s=\frac{1}{{\cal O}(1)}, N_f={\cal O}(1)$ (as in our MQGP limit), so that $\tilde r_0(B)=\tilde r_0(0)-|c_1|B$, i.e.\ $\gamma_{r_{h,0}}>0$ as quoted;
the on-shell value is
{\footnotesize
\begin{eqnarray}
\label{dr0B-onshell-SM-i}
\left.\partial_{\tilde r_0}{\cal B}\right|^{\rm on-shell}
&=&-3g_s^{3/2}\sqrt{\kappa_{\rm EH,IR}^{{\rm th},\beta}M}
 N^{1/20}N_f^{1/6}e^{\frac{6g_sN_f}{4\pi-g_sN_f\log2}+1}\nonumber\\[-1pt]
&&\times\sqrt{\frac{4\pi-g_sN_f(\log2-6)}
 {4\pi-g_sN_f\log2}}<0 .
\end{eqnarray}
}
Substituting (\ref{dr0B-onshell-SM-i}) into (\ref{rh-r0-beta0-SM}), one obtains:
\begin{equation}
\label{B-rh}
{\cal B}(\tilde r_h)=C_1-\tfrac{2{\cal C}_0}{3}\big[\log N-3\log\tilde r_h\big]^{-1/2}
\tilde r_h^{-3/2},
\end{equation}
with:
\begin{equation}
\label{constantC0-SM}
\scriptsize
\resizebox{\columnwidth}{!}{$\begin{aligned}
{\cal C}_0\sim{}&-
\frac{g_s^{7/8}
 M e^{-\frac{9g_sN_f}{4\pi-g_sN_f\log2}-\frac32}}
 {N_f^{1/3}\log^{1/3}N}\frac{\sqrt{6g_sN_f-g_sN_f\log2+4\pi}}
 {\sqrt{4\pi-g_sN_f\log2}}
\nonumber\\
&\times
\frac{\sqrt{18g_s^2N_f^2-3g_s^2N_f^2\log2
 +12\pi g_sN_f-4\pi g_sN_f\log2+16\pi^2}}
 {\sqrt{18g_s^2N_f^2+g_s^2N_f^2\log^22-g_s^2N_f^2\log8
 +12\pi g_sN_f-4\pi g_sN_f\log4+16\pi^2}};
\end{aligned}$}
\end{equation}
inversion yields the black-hole horizon radius as a function of the field,
\begin{equation}
\label{rh-B-expansion-SM}
\footnotesize
\begin{aligned}
\tilde r_h(B)&=\tilde r_h(0)-
\frac{B\,\tilde r_h(0)^{5/2}\sqrt{\log N-3\log\tilde r_h(0)}}
{{\cal C}_0[1-1/(\log N-3\log\tilde r_h(0))]},\\[-1pt]
\tilde r_h(0)&=\frac{(-1)^{1/3}(2/3)^{2/3}{\cal C}_0^{2/3}}
{C_1^{2/3}W[-4{\cal C}_0^2/(9C_1^2N)]^{1/3}} .
\end{aligned}
\end{equation}
with $W$ the Lambert function and ${\cal C}_0^2/C_1^2=9\,r_h(0)^3\log\!\big(N/r_h(0)^3\big)$.
Both branches thus give $\tilde r_{h,0}(B)=\tilde r_{h,0}(0)-\gamma_{r_{h,0}}B$
with $\gamma_{r_{h,0}}>0$, and the resulting horizon datum
$(\partial_{r}{\cal B})^2$ feeds \eqref{eq:DBI7}. Identifying $\epsilon_1^{3/2}\sim"e^{-t_\infty}", \ln N \sim "e^{t_c}"$ ($t_c$ being the IR cutoff) of \cite{Maldacena+Martelli}, interestingly, the ratio of the ${\cal O}(B)$ and ${\cal O}(B^2)$ coefficients in
\eqref{r0tilde-B-SM}, can be interpreted as the product of the cube roots respectively of the real and imaginary parts of the periods over the non-compact and compact three-cycles $\Gamma$ and $S^3$ of a deformed conifold, and is additionally large $N$-suppressed - see (\ref{OB-to-OBsq-ratio-beta0}). Numerically, this ratio works to about a little under $1\%$. Thus the monotonic, sign-definite response
$\partial{\cal B}(r_h)/\partial B_{\rm ext}>0$ is \emph{derived} here from
\eqref{DBI-selfcontained}--\eqref{rh-B-expansion-SM}, not posited; the companion
work \cite{Keshav:inprep} supplies only the lengthy intermediate algebra. 

The full DBI
expression for the ratio, whose on-shell value at
$Z{=}0\!\leftrightarrow\!r{=}r_{h/0}$ yields (in $\alpha'{=}1$ units, all
quantities dimensionless) the regime relations
\eqref{eq:regimes}, is ($G\equiv i^*(g+B_{\rm NS\text{-}NS})^{\rm IIA}$,
$i:\Sigma_7^{D7}\hookrightarrow M_{10}$)
\begin{equation}
\footnotesize
\resizebox{\columnwidth}{!}{$\begin{aligned}
\mathcal D_0 &\equiv
T^2N_fT_{\rm D6}
\frac{\dbimeas\!{\cal I}(\theta_2)
(\sqrt G G^{ZZ}G^{x^1x^1})^{\beta^0}(\partial_TB)^2}
{\dbiden}\,\dbiev,\\[-1pt]
\mathcal D_1 &\equiv
T^2N_fT_{\rm D6}
\frac{\dbimeas\!{\cal I}(\theta_2)
(\sqrt G G^{ZZ})^{\beta^0}(G^{x^1x^1})^{\beta}(\partial_TB)^2}
{\dbiden}\,\dbiev,\\[2pt]
\left.\frac{\zeta}{\eta}\right|_{\rm DBI}^{B}
&=\left[\tfrac19-\tfrac23(\tfrac13-c_s^2)^{\beta^0}\right]\mathcal D_0
\nonumber\\[-1pt]
&\quad+\beta\Big\{
\left[-\tfrac23(\tfrac13-c_s^2)^{\beta}
 +(\tfrac13-c_s^2)^{2,\beta}\right]\mathcal D_0
\nonumber\\[-1pt]
&\qquad\quad+
\left[\tfrac19-\tfrac23(\tfrac13-c_s^2)^{\beta^0}\right]\mathcal D_1
\Big\},\qquad T>T_c .
\end{aligned}$}
\label{eq:DBI7}
\end{equation}
${\cal I}(\theta_2)$ being an angular measure derived in \cite{Keshav:inprep}.
The $\beta^0$ terms are the leading two-derivative contributions and the
$\beta$ terms the ${\cal O}(R^4)$ corrections; the field enters through
$(\partial_r B)^2$ via the on-shell profile ${\cal B}(r)$, and is the origin of
the $(q_fB)^2$ dependence, hence of $\chi_B$.  The detailed KK reduction
generating ${\cal B}(r)$ and the full DBI evaluation \eqref{eq:DBI7} are given in
the companion works \cite{Symp-half-flat-SU3-struc,Kushwah2024,Keshav:inprep};
what is self-contained \emph{here}, and sufficient to verify the sign claim,
is the kinetic master identity \eqref{eq:master-SM}, its tabulated sign-definite
inputs (TABLE~\ref{tab:deconf}), and the lattice construction
\eqref{eq:phases}---the top-down input enters only as the sign of $\chi_B$,
which the lattice $c_s^2(B,T)$ independently fixes.

For physical $N_f=2{+}1$ quark masses, the deconfinement transition in QCD
is an analytic crossover rather than a genuine phase transition.  The
top-down dual computes the $N_f=3$ \emph{chiral-limit} idealization, where
the transition is first order (IMC-shifted Hawking--Page) at $T_0$ and
$\zeta/\eta$ is formally discontinuous there.  There is no tension with the
physical crossover: turning on physical quark masses smooths the first-order
jump (standard Columbia-plot expectation), and this is realized
\emph{explicitly} below---the $\zeta/\eta(T,B)$ we build from physical-mass
lattice $c_s^2(B,T)$ is continuous through $T_0$ (\LFmain,
dashed) while still exhibiting slope reversals on \emph{both} sides.  The
first-order dual enters only as a per-branch predictor of
$\mathrm{sgn}\,\partial_B(\zeta/\eta)$ away from $T_0$---a statement local to
$T\!\neq\!T_0$ that the smoothing leaves intact---and the physical-mass
construction joins the two branches continuously.  So the discontinuity is a
chiral-limit artifact, not a prediction for the physical theory, and the
double crossover survives the smoothing.  The odd
(parity-violating, Hall-viscosity-type) linear-in-$B$ terms---generated
in the dual by the parity-odd type-IIB axion (sourced by the $N_f$
flavor D7-branes) and the parity-odd, warp-factor-dependent type-IIB
ten-dimensional five-form, both entering the IIA mirror via SYZ triple
T-duality---are present but suppressed,
so the quadratic (parity-even) coefficient carries the magnetic structure
in each phase.  What will be shown in the next section is that even though $\zeta/\eta$ is continuous at the weak-magnetic-field shifted crossover transition temperature, the magnetic crossover predicted from the ${\cal M}$/string-theoretic dual, can also be realized from the available lattice data on $c_s^2$-vs-temperature for different magnetic fields.

\section{The two crossovers}
\label{sm:cross}

Equation~\eqref{eq:master-SM} is a competition between the source term~(I)
$=-4\chi_B/(\tfrac13-c_s^2)$ and the rate term~(II) $=-\Delta\gamma$.

\emph{Deconfined phase ($T>T_0$).}  Here the carriers are
asymptotically-free quarks scattering by $qq\to qq$ one-gluon exchange,
with internal-line (``loop'') mass
$m_{\rm loop,\ B=0}=\tfrac12 g(T)\sqrt{C_F}\,T$, $C_F=\tfrac43$.  The two
deconfined sub-regimes are set by whether $T$ lies below or above
$m_{\rm loop,\ B=0}$, i.e. by the sign of $\tfrac12 g(T)\sqrt{C_F}-1$, and
TABLE~\ref{tab:deconf} evaluates each term of \eqref{eq:master-SM} in both,
its bottom row assembling
$\partial\ln(\zeta/\eta)/\partial(q_fB)^2=({\rm I})-({\rm II})$,
$({\rm II})=\gamma_\zeta-\gamma_\eta$.  The
crossover is read off \emph{directly} from that row: for
$T_0<T<m_{\rm loop,\ B=0}$ ($\tfrac12 g\sqrt{C_F}>1$) one has, from
TABLE~\ref{tab:deconf}, $\gamma_\zeta=\tfrac{56}{3}A$ and
$\gamma_\eta=\tfrac{112}{5}A$ with $A=[(\tfrac14 g^2C_F)^{5/2}T^4]^{-1}>0$, so
that $({\rm II})=\gamma_\zeta-\gamma_\eta=(\tfrac{56}{3}-\tfrac{112}{5})A
=-\tfrac{56}{15}A<0$; hence $-({\rm II})>0$, and with
$({\rm I})\stackrel{<}{\sim}0$ the combination
$({\rm I})-({\rm II})>0$, giving $\partial_B(\zeta/\eta)>0$; for
$T>m_{\rm loop,\ B=0}$ ($\tfrac12 g\sqrt{C_F}<1$)
the running of $g(T)$ ($g'<0,g''>0$) makes both negative, $({\rm I})<0$ and
$({\rm II})<0$ with $-({\rm II})<0$, giving $\partial_B(\zeta/\eta)<0$.  The sign-definite jump
$\partial_B(\zeta/\eta):\,{>}0\to{<}0$ between the two columns is the upper
crossover, located at $T_2\sim m_{\rm loop,\ B=0}>T_0$, the $B=0$ internal
quark-loop scale.  Two caveats on this scale.  First, $T_2\sim
m_{\rm loop,\ B=0}$ (i.e.\ $\alpha\sim1$) is not \emph{imposed} to be the
crossover: the two columns are analytically distinct regimes with
\emph{opposite} slope signs, so by the intermediate-value theorem a zero of
$\partial_B(\zeta/\eta)$ must lie between them, near where they exchange
dominance, $\alpha\sim1$; the location follows from the sign flip, it is not
an assumption.  Second, the perturbative quasiparticle estimate is used only
for the \emph{scale and existence} of $T_2$, not a precision value---near
$150$--$200$~MeV weak coupling is not controlled, and the quoted
$T_2\simeq187$~MeV is read from the lattice construction
(\LFmain, $\xi_2{=}0$), with the kinetic estimate supplying only
that a reversal exists at ${\cal O}(m_{\rm loop})$ and its sign.  Because every
entry is a ratio of brackets on the same
single-mode Ansatz, the overall normalisation $\mathcal N_X$ cancels and
the \emph{existence} of $T_2$ is independent of it.  The confined-side
slope reversal $T_1<T_0$ requires a different computation, treated next.
The closed forms in TABLE~\ref{tab:deconf} follow from the
${\cal O}((q_fB)^2)$ quark thermal mass, whose field shift is set by the
dimensionless loop self-energy $X(\alpha)$ with column variable
$\alpha\equiv\tfrac12 g(T)\sqrt{C_F}$, $m_{\rm loop,\ B=0}=\alpha T$
\cite{Keshav:inprep}, where (with $s\equiv\sqrt{x^2+\alpha^2}$)
\begin{equation}
X(\alpha)=\!\int_0^\infty\!\! dx\,
\frac{s^2(e^{s}{+}1)+s^3 e^{s}}{2\,s^5(e^{s}{+}1)^2},
\label{eq:Xalpha}
\end{equation}
with $X\!\xrightarrow{\alpha\ll1}\!\tfrac{1}{4\alpha}$ and
$X\!\xrightarrow{\alpha\gg1}\!\tfrac{\sqrt\pi}{2^{3/2}}
e^{-\alpha}\alpha^{-3/2}(1{+}\alpha^{-1}{+}\cdots)$, the latter generating
the $e^{-\frac12 g\sqrt{C_F}}/(g\sqrt{C_F})^{3/2}$ factor of the
source-piece cell.  The rate log-derivatives $\gamma_\zeta,\gamma_\eta$
($>0$ on the space-like internal line, with $\ell_g\equiv\ln\alpha$) are
given in closed form in the corresponding rows of TABLE~\ref{tab:deconf},
completing its inputs.

\begin{table*}[t]
\renewcommand{\arraystretch}{1.7}
\setlength{\tabcolsep}{6pt}
\centering
\caption{Term-by-term evaluation of the deconfined master identity
\eqref{eq:master-SM} in the two sub-regimes $T\lessgtr m_{\rm loop,\ B=0}$
($m_{\rm loop,\ B=0}=\tfrac12 g(T)\sqrt{C_F}\,T$, $C_F=\tfrac43$, $N_c=3$).
The bottom row assembles
$\partial\ln(\zeta/\eta)/\partial(q_fB)^2=({\rm I})-({\rm II})$,
$({\rm II})=\gamma_\zeta-\gamma_\eta$, and shows the sign of
$\partial_B(\zeta/\eta)$ flipping from $>0$ to $<0$ across the columns:
this jump \emph{is} the upper crossover at $T_2\sim m_{\rm loop,\ B=0}>T_0$. Assuming $1/3 - c_s^2>0$, for $T<m_{\rm loop,\ B=0}, (I) - (II) \approx -(II) = \frac{3.73}{\big(\tfrac14 g^2 C_F\big)^{5/2}T^{4}}>0$; for $T>m_{\rm loop,\ B=0}$, and assuming $g'(T)<0, g''(T)>0$, $(I)<0$ and $-(II) = - \biggl(\dfrac{5}{\tfrac74\pi^2\big(\tfrac14 g^2 C_F\big)T^{4}} + \dfrac{15}{14\pi^4 T^{4}}\Big(3\ln\!\big(\tfrac12 g^2 C_F\big)+5.6\Big)\biggr)<0$.}
\label{tab:deconf}
\begin{ruledtabular}
\begin{tabular}{l c c}
 & $T<m_{\rm loop,\ B=0}\ \big(\tfrac12 g(T)\sqrt{C_F}>1\big)$
 & $T>m_{\rm loop,\ B=0}\ \big(\tfrac12 g(T)\sqrt{C_F}<1\big)$ \\[2pt]
\hline
(I)$\,=-\dfrac{4\,\partial_{(q_fB)^2}c_s^2}{\tfrac13-c_{s,0}^2}$
 & $\dfrac{-1}{\tfrac13-c_{s,0}^2}\cdot
    \dfrac{m_{\rm loop,\ B=0}}{T^{5}}\,
    \dfrac{e^{-\frac12 g\sqrt{C_F}}}{\big(g\sqrt{C_F}\big)^{3/2}}\ \lesssim 0$
 & $\dfrac{-1}{\tfrac13-c_{s,0}^2}\!\left[(g^2)'\ell_g-T(g^2)''\ell_g
    -2T(g')^2\right]<0$ \\
 & & $\ell_g\equiv\ln\!\big(\tfrac12 g\sqrt{C_F}\big),\quad
     (\because\, g'<0,\,g''>0)$ \\[2pt]
\hline
$\gamma_\zeta$
 & $\dfrac{56}{3}\,\dfrac{1}{\big(\tfrac14 g^2 C_F\big)^{5/2}T^{4}}$
 & $\dfrac{5}{\tfrac74\pi^2\big(\tfrac14 g^2 C_F\big)T^{4}}$ \\[2pt]
\hline
$\gamma_\eta$
 & $\dfrac{112}{5}\,\dfrac{1}{\big(\tfrac14 g^2 C_F\big)^{5/2}T^{4}}$
 & $\dfrac{-15}{14\pi^4 T^{4}}\Big(3\ln\!\big(\tfrac12 g^2 C_F\big)+5.6\Big)
    $ \\[2pt]
\hline
$\dfrac{\partial\ln(\zeta/\eta)}{\partial(q_fB)^2}=({\rm I})-({\rm II})$
 & $>0$
 & $ <0$ \\
 & $\Longrightarrow\ \partial_B(\zeta/\eta)>0$
 & $\Longrightarrow\ \partial_B(\zeta/\eta)<0$ \\
\multicolumn{3}{c}{\textbf{crossover:}\ \ 
 $\partial_B(\zeta/\eta)\big|_{T<m_{\rm loop,\ B=0}}>0
  \ \xrightarrow{\ T=T_2\,\sim\, m_{\rm loop,\ B=0}\ }\ 
  \partial_B(\zeta/\eta)\big|_{T>m_{\rm loop,\ B=0}}<0$} \\
\end{tabular}
\end{ruledtabular}
\end{table*}

\section{The confined phase: a chiral pion gas}
\label{sm:pion}

A caveat on the scope of \eqref{eq:master-SM} is in order, since no single
microphysical computation is controlled across the whole phase diagram: a
weakly coupled quark--gluon kinetic theory, a small-$(q_fB)$ expansion and
a single-mode ansatz cannot all hold below $T_0$, where the carriers are
hadrons.  What is universal in \eqref{eq:master-SM} is its \emph{form}, not
its inputs: for any $X=\mathcal N_X^2/\Gamma_X$ the ratio slope is fixed
algebraically by the source and rate log-derivatives,
$2(\nu_\zeta-\nu_\eta)-\Delta\gamma$, independent of which quasiparticle
realises $\mathcal N_X,\Gamma_X$.  The microphysics enters only through
those two, which we evaluate in the \emph{phase-appropriate} truncation on
each side---$qq\to qq$ one-gluon exchange for $T>T_0$
(TABLE~\ref{tab:deconf}) and the chiral pion gas below.  So
\eqref{eq:master-SM} is an identity within a specified truncation, applied
piecewise, not one weak-coupling formula extrapolated through $T_c$; the
near-$T_0$ window, where neither truncation is parametrically clean, is
precisely where the two are matched by continuity (\LFmain)
rather than by a single calculation.

Below $T_0$ the carriers are the
(nearly) massless pseudo-Goldstone bosons of $SU(3)_L\times SU(3)_R\to
SU(3)_V$; the field enters at leading order through the charged pion
$t$-channel exchange ($\pi^\pm$), whose squared mass takes the
lowest-Landau-level value $m^2(|eB|)=m^2+|eB|+O((eB)^2)$
\cite{Weinberg66,Bali18,Ding21}.  In the exact three-flavor chiral limit the
charged kaons $K^\pm$ are also massless charged Goldstone modes and receive the
same LLL shift; we retain only the charged-pion channel because at physical
strange-quark mass the kaon contribution to the transport integrals is
Boltzmann-suppressed by $e^{-m_K/T}$ relative to the pion one throughout the
confined window $T<T_0\simeq143$~MeV ($m_K/T_0\gtrsim3.4$, so $e^{-m_K/T_0}\lesssim0.03$), and the confined-side result is a
leading-order \emph{sign} statement.  The chiral-limit bookkeeping is thus used
only where the pion channel dominates; restoring $K^\pm$ shifts magnitudes, not
the endpoint signs, at the $O(e^{-m_K/T})$ level.
The
amplitude correction
\begin{equation}
|\mathcal M|^2=|\mathcal M|^2_0\Big[1-\tfrac{|eB|}{m^2-q^2}\,w
+O\big((eB)^2\big)\Big],\quad w>0,
\label{eq:Mpion}
\end{equation}
has $w>0$ on the space-like exchange.  The chiral
normalisation $|\mathcal M|^2_0$ cancels in each
$\partial_{|eB|}\ln\zeta,\partial_{|eB|}\ln\eta$ separately, so the
slope-reversal condition is independent of the overall interaction strength,
and the soft sector is infrared finite even for massless Goldstone
carriers.  This fixes the leading-${\cal O}(|eB|)$ \emph{sign} of the ratio
slope: the external charged-leg Landau quantisation and phase-space effects
are also ${\cal O}(|eB|)$ but common-mode to the bulk and shear collision integrals,
dividing out of $\zeta/\eta$ and leaving the exchanged-line shift as the
surviving tensor-distinguishing effect; a full magnetised pion-gas transport
computation (external Landau levels, modified phase space, field-dependent
projectors) is not carried out here or in the companion, so the confined-side
result is a leading-${\cal O}(|eB|)$ \emph{sign} statement, not a complete
calculation.  The master
expression retains the form \eqref{eq:master-SM}
with $\widetilde\chi\equiv\partial c_s^2/\partial|eB|$ replacing $\chi_B$.

\emph{The high-$T$ endpoint at the crossover.}  For physical $N_f=2{+}1$
quark masses, the transition is an analytic crossover, not a first-order
jump: there is no mixed phase and $c_s^2$ does not vanish but
passes through a \emph{finite minimum} near $T_0$ (lattice and $\chi$PT
both give $c_s^2(T_0^-)\simeq0.15$, i.e. $\tfrac13-c_s^2\simeq0.18$),
\begin{equation}
\tfrac13-c_s^2\big|_{T\to T_0^-}\to u_{\min}\simeq0.18,\qquad
\frac{1}{\tfrac13-c_s^2}\Big|_{T_0^-}\simeq5.5,
\label{eq:cbendpoint}
\end{equation}
a finite, modest amplification at the upper end.  The low-$T$ end is fixed
by the chiral pion gas: in the chiral limit at vanishing isospin chemical
potential the three-loop $\chi$PT pressure is $p(T)=A(T)\,T^4$,
$A(T)=\tfrac1{30}\pi^2\big[1+\tfrac{T^4}{36F^4}\ln\tfrac{\Lambda_p}{T}\big]$
\cite{Gerber89}.  Then $c_s^2=p'(T)/[T\,p''(T)]$, and for $T\ll\Lambda_p$
yields $\tfrac13-c_s^2\sim\tfrac{T^4}{F^4}\ln\tfrac{\Lambda_p}{T}$, which is
\emph{positive} in the entire confined window
$T<T_0\ll\Lambda_p\simeq275$~MeV \cite{Gerber89}:
\begin{equation}
c_s^2(T)\xrightarrow{T\to0^+}\tfrac13^{-},\qquad
\frac{1}{\tfrac13-c_s^2}\xrightarrow{T\to0^+}\gg1 .
\label{eq:loTendpoint}
\end{equation}
The $c_s$-dependent term on the RHS of (\ref{eq:master-SM}) is therefore
positive and large at the low-$T$ end, but much smaller (though finite)
near $T_0^-$; this asymmetry drives the confined-side slope reversal.  Two
scope points.  \eqref{eq:master-SM} is a \emph{logarithmic} derivative: as
$\zeta/\eta\sim u^2\to0$ ($u\to0$) its low-$T$ divergence is offset by a
vanishing prefactor, so the absolute slope
$\partial_{|eB|}(\zeta/\eta)=(\zeta/\eta)\,\partial_{|eB|}\ln(\zeta/\eta)$
stays bounded---we claim only the \emph{sign} and the reversal \emph{location}
$T_1$, not a large low-$T$ magnitude; and since $\zeta/\eta>0$ on $(0,T_0)$,
$\mathrm{sgn}\,\partial_{|eB|}(\zeta/\eta)=\mathrm{sgn}\,\partial_{|eB|}
\ln(\zeta/\eta)$, so $T_1$ is unaffected.  The chiral limit
($c_s^2\to\tfrac13^-$) is used only to fix $\mathrm{sgn}(u)>0$ on $(0,T_0)$
(robust to ${\cal O}(m_\pi^2/T^2)$); all magnitudes ($u_{\min}$, $\widetilde\chi$,
$T_0$) and the crossover continuity are physical-mass inputs.

\emph{The source is the chiral susceptibility, not GMOR.}  In the chiral
limit $m_q\langle\bar qq\rangle=-m_\pi^2 f_\pi^2\to0$, so the
explicit-breaking (Gell-Mann--Oakes--Renner) term of the trace anomaly
vanishes and cannot carry the field response.  Instead, by the chain rule,
$\widetilde\chi=(\partial c_s^2/\partial\langle\bar qq\rangle)
(\partial\langle\bar qq\rangle/\partial|eB|)$, with both factors finite on
$(0,T_0)$.  We use $\widetilde\chi<0$ (hadronic inverse magnetic catalysis)
\cite{Goswamietal23,Bali14} only in the near-crossover region $T\to T_0^-$,
where it is a robust lattice statement; we do \emph{not} assume it holds down to
$T=0$ (at low $T$ ordinary magnetic catalysis is expected, and the sign of
$\widetilde\chi$ may reverse).  The low-$T$ endpoint of the slope does not rely
on $\widetilde\chi$: it is set by the lowest-Landau-level charged-pion mass shift
$m^2(|eB|)=m^2+|eB|$, which raises the exchanged-pion mass and, through
\eqref{eq:Mpion} with $w>0$, fixes $\partial_{|eB|}(\zeta/\eta)>0$ as
$T\to0^+$ independently of the condensate response.  The confined-phase source term is
$\mathcal S=-4\widetilde\chi/(\tfrac13-c_s^2)$ with $\widetilde\chi<0$.
By the endpoints \eqref{eq:cbendpoint}--\eqref{eq:loTendpoint},
$\mathcal S\gg1$ as $T\to0^+$ (giving $\partial_{|eB|}(\zeta/\eta)>0$)
while $\mathcal S|_{T_0^-}=-4\widetilde\chi(T_0^-)/u_{\min}$ is finite; as
the bounded rate term ($\gamma_\zeta,\gamma_\eta>0$, finite) can then
overturn the sign near $T_0^-$, a slope reversal $T_1<T_0$ exists
\emph{whenever}
\begin{equation}
\frac{4\,|\widetilde\chi(T_0^-)|}{u_{\min}}<\Delta\Gamma_0,
\label{eq:straddle}
\end{equation}
a single finite condition at the upper end---sufficient, not a proof of
$T_1$ from a full chiral kinetic calculation: it states that a reversal
occurs \emph{if} the finite near-$T_0$ rate contribution exceeds the finite
source contribution, which the deconfined rate growth and the sign inputs
make natural.  We are explicit that this inequality is an
\emph{assumption} we do not prove: a first-principles magnetised pion-gas
transport computation establishing \eqref{eq:straddle} lies beyond both this
Letter and its companion, and the confined-side reversal $T_1$ is therefore
reported as \emph{conditional} on it throughout.
Since $\varepsilon$,
$\langle\bar qq\rangle$ and the screening mass vary smoothly through $T_0$
in the physical-mass crossover, $\partial_{|eB|}(\zeta/\eta)$ is
continuous across $T_0$, matching the continuity of $\zeta/\eta$ in
\eqref{eq:phases} and \LFmain.

\begin{figure}[t]
\centering
\includegraphics[width=0.94\columnwidth]{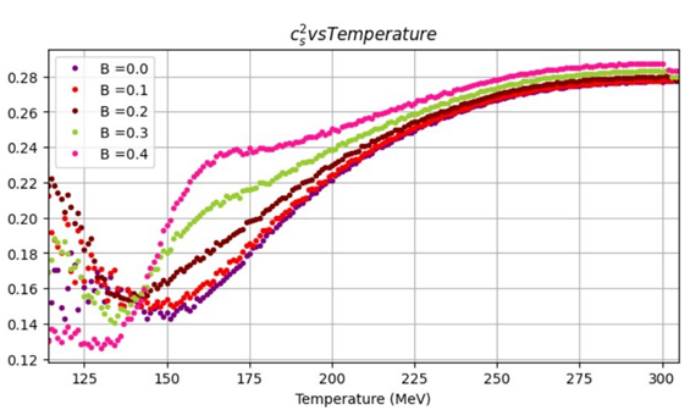}
\caption{Input lattice $c_s^2$ vs $T$ for $B=0$--$0.4$ \cite{Bali14}.
The minimum shifts to lower $T$ with increasing $B$, and
$\partial_B c_s^2$ reverses sign across $T_0\simeq143$~MeV
(hadronic inverse magnetic catalysis below, ordinary above), the input
feeding the crossover construction \eqref{eq:phases}.}
\label{fig:cs2input}
\end{figure}



\section{Lattice construction across the crossover}
\label{sm:lat}

Bali \emph{et~al.}~\cite{Bali14} provide $c_s^2(B,T)$ but not
$\zeta/\eta$ directly; their simulations use \emph{physical} $N_f=2{+}1$
quark masses, for which the thermal transition is an analytic crossover
with a pseudo-critical temperature lowered by the field (inverse magnetic
catalysis).  The pressure--energy data sets at different $B$, given in
\cite{Bali14}, were used to compute $c_s^2$ numerically (FIG.~\ref{fig:cs2input})
and hence $\zeta/\eta$ (\LFmain).

With $u(T,B)\equiv\tfrac13-c_s^2(T,B)>0$, and motivated by the fact that in
two-body--scattering kinetic theory
$\zeta(T,B)=\int\frac{d^3p}{(2\pi)^3}\,\tau(T,B,|\vec p\,|)\,
\mathcal D^2(\vec p,T,B)\,\frac{f^{(0)}(1-f^{(0)})}{E_p}$,
$E_p=\sqrt{\vec p^{\,2}+m^2}$, with bulk driving term
$\mathcal D=[\,\vec p^{\,2}/3E_p^2-c_s^2(T,B)]E_p$,
$\mathcal D^2=\mathcal D_0^2-2\mathcal D_0(q_fB)^2\Phi+O((q_fB)^4)$, and inverse
relaxation time $\tau^{-1}=-\,\mathrm{Im}\,\Sigma^{(2)}_R(E_p,\vec p;B,T)/E_p$,
the field enters at ${\cal O}((q_fB)^2)$.  We expand $\zeta/\eta$ in the natural
$d$-quark field variable $|q_d|B=B/3$ ($q_d=-\tfrac13$), per phase,
[convention: the lattice quotes $eB$; the kinetic identity is written
per-flavor in $(q_fB)^2$ with $q_f$ the charge of the exchanged quark, and
we set the expansion scale by the down quark, $x\equiv|q_d|B=B/3$, since it
dominates the confined-side charged-pion channel and carries the smallest
charge (most conservative).  The charge weights $q_{u,d,s}=(\tfrac23,
-\tfrac13,-\tfrac13)e$ enter only the \emph{magnitudes} $\xi_n$, as an
overall charge-weighted sum $\sum_f q_f^2$; the \emph{sign} of
$\partial_B(\zeta/\eta)$---the sole claim here---is invariant under the
choice of charge variable, since rescaling $B\to q_fB$ is a positive
monotone map that cannot flip a slope's sign.]
\begin{equation}
\frac\zeta\eta(T,B)=
\begin{cases}
\sum_{n=0}^{2}\widetilde\xi_n(T)\,(|q_d|B)^n, & T<T_0,\\[2pt]
\sum_{n=0}^{2}\xi_n(T)\,(|q_d|B)^n, & T>T_0,
\end{cases}
\label{eq:phases}
\end{equation}
with the continuity matching $\zeta/\eta(T_0^-,B)=\zeta/\eta(T_0^+,B)$ and
$|\xi_1/\xi_2|,|\widetilde\xi_1/\widetilde\xi_2|=O(0.01)$.  The two
measured fields, $x\equiv|q_d|B=0.1$ ($B=0.3$) and $x=0.1\overline3$
($B=0.4$), give the determination equations
\begin{align}
\frac\zeta\eta(T,B{=}0.3)&=\zeta_0+\zeta_1(0.1)+\zeta_2(0.1)^2,\label{eq:qdBdet1}\\
\frac\zeta\eta(T,B{=}0.4)&=\zeta_0+\zeta_1(0.1\overline3)+\zeta_2(0.1\overline3)^2,\label{eq:qdBdet2}
\end{align}
with $(\zeta_n)=(\widetilde\xi_n)$ for $T<T_0$ and $(\xi_n)$ for $T>T_0$.
The coefficient $\zeta_1$ here is \emph{not} a free closure parameter of the
parity-even scalar: it is the amplitude of the separately-projected parity-odd
structure $\mathcal R^{\rm odd}_1$ (a $|q_dB|$ threshold piece, distinct in
tensor character from the even $\mathcal R$; see ``The separate parity-odd (Hall)
sector'' below), and its ratio to $\zeta_2$ is \emph{fixed} by the top-down
dual's odd/even ratio $|\mathcal R^{\rm odd}_1/\mathcal R_2|\sim10^{-2}$, not
tuned to the data.  With that ratio taken as an input, the two measured fields
determine $\zeta_0,\zeta_1,\zeta_2$
uniquely, reproducing the measured curves exactly, so positivity and the
four-region structure are inherited.  Since $|\zeta_1|\ll|\zeta_2|$,
$\partial_B(\zeta/\eta)\propto\zeta_2(T)$ at the data fields, and the
zeros of $\zeta_2$ locate the crossovers \emph{within the imposed $\zeta/\eta$
ansatz and the chosen odd/even closure}.
We stress the status: the ansatz, the closure $\zeta_1=0.01\zeta_2$, and the
coefficient choices below are imposed, so this construction does not
\emph{derive} the sign change from lattice QCD---it exhibits that the
available lattice $c_s^2(B)$ is \emph{consistent with} a $\zeta/\eta$ whose
magnetic slope reverses; the \emph{existence} of the reversals rests on the
kinetic identity and the string dual, not on this fit.

These coefficients are generated from the $u=\tfrac13-c_s^2$ ansatz
$\zeta/\eta=a_1u+a_2u^2$ ($T<T_0$), $a_3u+a_4u^2$ ($T>T_0$), with
$B$-dependent $a_i(T,B)=a_{i1}+a_{i2}B+a_{i3}B^2$.  Fixing $a_{41}=a_{21}=1.2$ (enforcing $a_4,a_2$-dominance) and the $u^2$ slot
$B$-independent, and feeding in $\chi=\partial_Bc_s^2$,
$\kappa=\tfrac12\partial_B^2c_s^2$ from a quadratic fit of the five-field
lattice $c_s^2(B)$ ($B=0$--$0.4$) at each $T$, the $\xi_n,\widetilde\xi_n$
(and their inverse maps to $a_{ij}$) follow in closed form
\cite{Keshav:inprep}.  The construction imposes $a_1>0$, $a_4>0$;
magnitude dominance $|a_2|<a_1$ ($T\!\ll\!T_0$), $|a_3|<a_4$
($T\!\gg\!T_0$); $\zeta/\eta>0$; and $\xi_n\neq\widetilde\xi_n$ (at the
boundary $\xi_2>0$ while $\widetilde\xi_2<0$).  The physical
$\xi_n,\widetilde\xi_n$ are gauge-invariant though the $a_{ij}$ are not.
Using $B=0.3,0.4$ with the $<\!1\%$ Hall ratio as per-phase closure fixes
all coefficients, generating the four-region pattern
$\partial_B(\zeta/\eta)>0\ (T<T_1)$, $<0\ (T_1<T<T_0)$, $>0\ (T_0<T<T_2)$,
$<0\ (T>T_2)$, with $T_1<T_0<T_2$,
$T_1\simeq134~\mathrm{MeV}$ ($\widetilde\xi_2=0$),
$T_0\simeq143~\mathrm{MeV}$, $T_2\simeq187~\mathrm{MeV}$ ($\xi_2=0$),
continuous at $T_0$ (\LFmain).

\emph{Reproducibility of the numerics.}  The construction is fully specified by the following: (i)~$c_s^2(T,B)$ is obtained from the Bali \emph{et al.}\
pressure/energy tables \cite{Bali14} by finite differencing $dp/d\varepsilon$;
(ii)~at each $T$ a quadratic least-squares fit of $c_s^2$ in $B$ over
$B=0$--$0.4$ returns $(\chi_B,\kappa_B)$; (iii)~$\xi_n,\widetilde\xi_n$ follow
from \eqref{eq:phases}--\eqref{eq:qdBdet2} in closed form with the fixed
$<\!1\%$ odd/even closure; (iv)~$\partial_B(\zeta/\eta)\propto\xi_2,\widetilde
\xi_2$ and its zeros give $T_1,T_2$. 

\emph{The separate parity-odd (Hall) sector.}  The parity-odd Hall structures admitted by magnetized
constitutive relations~\cite{HK2017}, as mentioned in \ref{intro}, are activated here by the genuinely
parity-odd data of the dual---the type-IIB axion and the warped five-form (see
below)---whose presence, together with $B$, is what permits a Hall projection
in addition to the parity-even viscosities.  In our weak-field
window $|q_f|B\sim T_0^2$, this appears as the distinct projected quantity
$\mathcal R^{\rm odd}_1 B$ of \eqref{eq:Rdef-SM}---a lower-order ${\cal O}(B^1)$
tensor structure orthogonal to (not a correction of) the parity-even
$\mathcal R$ tracked by the identity \eqref{eq:master-SM}.  It is exactly the
linear-in-$B$ term predicted by the string dual (parity-odd type-IIB axion
and warped five-form), and in the fit it sits at
$|\mathcal R^{\rm odd}_1/\mathcal R_2|=|\xi_1/\xi_2|\sim10^{-2}$: present,
reported, but not part of the even-in-$B$ scalar whose slope reverses.
We emphasize the epistemic status sharply: lattice equation-of-state data are
\emph{parity-even} thermodynamic data, so they can test compatibility with the
even sector $\mathcal R_{\rm even}=\mathcal R_0+\mathcal R_2 B^2+\cdots$ but
\emph{cannot} validate the parity-odd projection
$\mathcal R_{\rm odd}=\mathcal R^{\rm odd}_1 B+\cdots$, which belongs to a
different Kubo projection.  The percent-level odd/even ratio is therefore a
\emph{top-down input}, not a quantity supported by lattice thermodynamics.

This shows lattice $c_s^2(B)$ can be made to be compatible with the top-down
results [(\ref{zetaoveretaB0})+(\ref{eq:DBI7})]; the \emph{existence}
of the crossovers can be explained by a kinetic identity \eqref{eq:master-SM}
(deconfined) and its pion-gas counterpart (confined), conditional on the
rate growth below.


\section{Mutual consistency of the frameworks}
\label{sm:mutual}

The crossover is corroborated across three approaches with no common
dynamical assumptions, with a fourth entering as a compatibility check on the
trend.  (i) The top-down string/$\mathcal M$-theory dual
predicts $\partial_B(\zeta/\eta)|_T>0\to<0$ on both sides of $T_0$
(FIG. 1 and FIG. 2 of the Letter).  (ii) In kinetic theory the
deconfined reversal at $T_2$ follows from the identity \eqref{eq:master-SM}
with tabulated, sign-definite inputs (TABLE~\ref{tab:deconf}), while the
confined reversal at $T_1$ follows under the sufficient condition
\eqref{eq:straddle} on the pion-gas source and rate---sign-consistent, not
a first-principles proof of $T_1$.  (iii) The lattice construction
\eqref{eq:phases} exhibits that physical-mass $c_s^2(B,T)$ is
\emph{consistent with} the same continuous, double-crossover structure
(\LFmain)---a construction compatible with the string
prediction, not an independent measurement of $\zeta/\eta$.  (iv)
The bottom-up V-QCD model \cite{Demircik2023,Demircik2024} provides an
independent holographic window at weaker coupling.  Digitizing its
anisotropic viscosities [FIG.~4.(a) of the Letter] gives
$\partial_B\mathcal R_\parallel<0$ throughout its published window
$1.08\le T/T_c\le1.80$, with \emph{no} sign change there; what the data do
show is that $|\partial_B\mathcal R_\parallel|$ decreases monotonically with
$T$ (from $-4.4\sigma$ at $1.08\,T_c$ to $-0.3\sigma$ at $1.80\,T_c$),
i.e.\ the slope approaches zero \emph{from below}---compatible with an upper
reversal $T_2$ lying just above the plotted range, but not a demonstration
of one.  We stress that the fine structure at small $eB$ is
\emph{sub-resolution}: the $eB=0.1$ and $0.2$ curves are separated by
$\lesssim2$ pixels of the source figure, below its linewidth [grey band in
FIG.~4.(b) of the Letter], so a crossing between them---such as the apparent
one near $1.425\,T_c$---can be neither confirmed nor excluded from the
published plot.
V-QCD therefore enters here as a
\emph{compatibility check on the trend}, not as a fourth confirmation of the
crossover; settling it requires the underlying numerical data, which their
closed-form horizon formulae supply directly.
That a UV-complete eleven-dimensional construction, a
flat-space collision integral, and lattice $c_s^2(B)$ \emph{converge} on one
crossover structure across the
strong-to-weak-coupling axis---each with independent systematics---is the
central result: not a single closed-form proof, but a mutually reinforcing
prediction that a direct determination of $\zeta/\eta(B,T)$ can now test.

\paragraph{\bf Acknowledgments.}
This work is supported in part by the Natural Sciences and Engineering Research Council of Canada (SAPIN-2025-00041 [KD], SAPIN-2026-00047 [CG], SAPIN-2024-00026 [SJ]). 
SSK is supported by a Senior Research Fellowship (SRF) from the Ministry of Education (formerly MHRD), Government of India. 
AM is partly supported by the Professional Development Allowance and the Professional Development Funds available from IIT Roorkee; he would also wish to thank the department of Physics, McGill University (KD, in particular) for the wonderful hospitality during his visits to McGill. 
FIG.~S1 was obtained by a Python code written by Open AI (ChatGPT) [that used the EoS data files of \cite{Bali14}], and compiled by Google collab. The numerical values of $\zeta_{0,1,2}, \tilde{\zeta}_{0,1,2}, a_{iI}, i=1,...,4; I=1,2,3$ in \ref{sm:lat} (and hence FIGS.~3, 4 of the main text), the kinetic theory identity (\ref{eq:master}), the expressions for the shape functions (\ref{eq:Gzeta-Geta}), the idea to look at $\delta g_{x^3x^3}/g_{x^3x^3}=(\Delta_L^{-1}T^{D6,\ell=2})_{x^3x^3}/({\cal G}^{-1}T^{\rm SUGRA})_{x^3x^3}$ (\ref{fracmetmaganisotropy}) to explicitly show the suppression of anisotropy, to look at (\ref{Wald_contribution}) to show the suppression of the contribution of the $J_0$ terms to $\frac{\zeta}{\eta}$, the expression for the quark-loop integral (\ref{eq:Xalpha}) and the chiral pion-gas discussion in the confined phase, were developed with the assistance of Claude AI Opus 4.8 (Anthropic). The manuscript has been prepared with great care by the authors, who assume full responsibility for its contents, including any errors or omissions that may have inadvertently remained.


\begin{thebibliography}{99}

\bibitem{HK2017} J.~Hernandez and P.~Kovtun,
\textit{Relativistic magnetohydrodynamics},
JHEP \textbf{05} (2017) 001 [arXiv:1703.08757 [hep-th]].

\bibitem{SM} See Supplemental Material for
the observable definitions and weak-field power counting (Sec.~I), the
derivation of the kinetic-theory identity (Sec.~II), the $\mathcal M$-theory
construction and the $O(R^4)$ analysis (Sec.~III), the confined-phase
pion-gas endpoints (Sec.~IV), and the lattice construction (Sec.~V), which
includes Refs.~\cite{MQGP,Kushwah2024,Hattori2017,Bali14,Goswamietal23,Weinberg66,%
Bali18,Ding21,Demircik2023,Demircik2024,MisraYadav2020,OnShellR4,Keshav:inprep}.

\bibitem{Maldacena+Martelli} J.~M.~Maldacena and D.~Martelli, ``The unwarped, resolved, deformed conifold: fivebranes and the baryonic branch of the Klebanov-Strassler theory,' JHEP \textbf{01} (2010) 104, arXiv:0906.0591 [hep-th].

\bibitem{Cg}The angular regularization factor ${\cal C}_g$, used in \cite{Kushwah2024}, is defined as:
{\footnotesize
\begin{eqnarray*}
\label{regularizationbeta}
& & \frac{1}{ {\epsilon_1}^9 {\epsilon_2}^4}\Big\{-3640 {\epsilon_1}^4 {\epsilon_2}^2
   ({\cal C}_{\theta_1z}-{\cal C}_{yz})\nonumber\\
& & -1960 {\epsilon_1}^2 {\epsilon_2}^4
   ({\cal C}_{\theta_1z}-{\cal C}_{yz}) \log ({\epsilon_2})\nonumber\\
& & +9 {\epsilon_1}^6 (272 {\cal C}_{\theta_1z}-210 {\cal C}_{yz})\nonumber\\
& & -237452 {\cal C}_{\theta_1z} {\epsilon_2}^4\Big\} ={\cal C}_g\equiv{\rm finite},
\end{eqnarray*}
}
${\cal C}_{MN}$ being constants of integration appearing in the solutions to the EOMs for the ${\cal O}(R^4)$ metric corrections and $\epsilon_{1,2}$ are angular regulators: $\theta_{1,2}\rightarrow0$ respectively as $\epsilon_{1,2}\rightarrow0$; $(x,y,z)$ being the toroial analogs of $(\phi_1,\phi_2,\psi)$ for the construction  of delocalized SYZ type IIA mirror. ${\cal C}_g=0$ implies that we choose a regularization wherein 
{\footnotesize
\begin{equation*}
\begin{aligned}
~~~&\epsilon_1   = \frac{\sqrt[6]{59363}\,{\epsilon_2}^{2/3}}{\sqrt[3]{3}\,\sqrt{136{\cal C}_{\theta_1z}-105{\cal C}_{yz}}}
\Big[-28560\,{\cal C}_{\theta_1z}^2{\cal C}_{yz}\\
&+|{\cal C}_{\theta_1z}|(136{\cal C}_{\theta_1z}-105{\cal C}_{yz})^2
+18496\,{\cal C}_{\theta_1z}^3+11025\,{\cal C}_{\theta_1z}{\cal C}_{yz}^2\Big]^{1/6}.
\end{aligned}
\end{equation*}
}

\bibitem{Kushwah2024} S.~S.~Kushwah and A.~Misra,
\textit{Bulk viscosity, speed of sound and contact structure at intermediate
coupling}, Phys. Rev. D \textbf{110}, no.12, 126010 (2024)
[arXiv:2403.10541 [hep-th]].

\bibitem{FEP} M.~Mia, K.~Dasgupta, C.~Gale and S.~Jeon,
\textit{Five Easy Pieces: The Dynamics of Quarks in Strongly Coupled Plasmas},
Nucl.\ Phys.\ B \textbf{839} (2010) 187
[arXiv:0902.1540 [hep-th]] ;
\textit{Toward Large $N$ Thermal QCD from Dual Gravity: The Heavy Quarkonium
Potential}, Phys.\ Rev.\ D \textbf{82} (2010) 026004 [arXiv:1004.0387 [hep-th]].

\bibitem{MQGP} M.~Dhuria and A.~Misra,
\textit{Towards MQGP}, JHEP \textbf{11} (2013) 001
[arXiv:1306.4339 [hep-th]].



\bibitem{MQGP-glueball-spectroscopy} K.~Sil, V.~Yadav and A.~Misra,
{\it Top{\textendash}down holographic G-structure glueball spectroscopy at (N)LO in $N$ and finite coupling}, Eur. Phys. J. C \textbf{77}, no.6, 381 (2017) [arXiv:1703.01306 [hep-th]].

\bibitem{MQGP-meson-spectroscopy} V.~Yadav, A.~Misra and K.~Sil,
{\it Delocalized SYZ Mirrors and Confronting Top-Down $SU(3)$-Structure Holographic Meson Masses at Finite $g$ and $N_c$ with P(article) D(ata) G(roup) Values},
Eur. Phys. J. C \textbf{77}, no.10, 656 (2017) [arXiv:1707.02818 [hep-th]].

\bibitem{MQGP-trace-anomaly} A.~Misra and C.~Gale, {\it The QCD trace anomaly at strong coupling from M-theory}, Eur. Phys. J. C \textbf{80}, no.7, 620 (2020) [arXiv:1909.04062 [hep-th]].

\bibitem{IITR-McGill-Bulk}A.~Czajka, K.~Dasgupta, C.~Gale, S.~Jeon, A.~Misra, M.~Richard and K.~Sil, {\it Bulk Viscosity at Extreme Limits: From Kinetic Theory to Strings},
JHEP \textbf{07}, 145 (2019) [arXiv:1807.04713 [hep-th]]; 
\textit{On bulk viscosity at weak and strong 't Hooft couplings},
Mod. Phys. Lett. A \textbf{35}, no.27, 2030012 (2020)
[arXiv:1807.07950 [hep-th]].

\bibitem{MQGP-transport} S.~S.~Kushwah and A.~Misra,
{\it $T_c$, Photoproduction, Paramagnetic Anisotropic Plasma, IR Log-Gravitational-DBI Renormalization, and $G_2$-Structure Induced (Almost) Contact 3-Structures in Hot Strongly Magnetic MQCD at Intermediate Coupling}, Fortsch. Phys. \textbf{73}, no.12, e70044 (2025)
[arXiv:2503.07732 [hep-th]];  K.~Sil and A.~Misra,
{\it New Insights into Properties of Large-N Holographic Thermal QCD at Finite Gauge Coupling at (the Non-Conformal/Next-to) Leading Order in N}, Eur. Phys. J. C \textbf{76}, no.11, 618 (2016) [arXiv:1606.04949 [hep-th]]; M.~Dhuria and A.~Misra,
{\it Transport Coefficients of Black MQGP M3-Branes},
Eur. Phys. J. C \textbf{75}, no.1, 16 (2015) [arXiv:1406.6076 [hep-th]].

\bibitem{ElingOz2011} C.~Eling and Y.~Oz,
JHEP \textbf{06} (2011) 007 [arXiv:1103.1657 [hep-th]].

\bibitem{Bali14} G.~S.~Bali, F.~Bruckmann, G.~Endrodi, S.~D.~Katz and
A.~Schafer, \textit{The QCD equation of state in background magnetic fields},
JHEP \textbf{08} (2014) 177 [arXiv:1406.0269 [hep-lat].

\bibitem{Demircik2024}  T.~Demircik, D.~Gallegos, U.~G\"ursoy,
M.~J\"arvinen and R.~Lier,
Phys.\ Rev.\ D \textbf{110} (2024) 066007
[arXiv:2402.12224 [hep-th]].

\bibitem{Hattori2017} K.~Hattori, X.-G.~Huang, D.~H.~Rischke and D.~Satow,
\textit{Bulk Viscosity of Quark-Gluon Plasma in Strong Magnetic Fields},
Phys.\ Rev.\ D \textbf{96} (2017) 094009 [arXiv:1708.00515 [hep-ph]].

\bibitem{Goswamietal23} K.~Goswami, D.~Sahu, J.~Dey, R.~Sahoo and R.~Stock,
\textit{Anisotropy of magnetized quark matter},
Phys.\ Rev.\ D \textbf{109} (2024) 074012 [arXiv:2310.02711 [hep-ph]].

\bibitem{Weinberg66} S.~Weinberg,
\textit{Pion scattering lengths}, Phys.\ Rev.\ Lett.\ \textbf{17} (1966) 616.

\bibitem{Bali18} G.~S.~Bali, B.~B.~Brandt, G.~Endr\H{o}di and
B.~Gl\"a\ss le, Phys.\ Rev.\ D \textbf{97} (2018) 034505
[arXiv:1707.05600 [hep-lat]].

\bibitem{Ding21} H.-T.~Ding \emph{et al.},
Phys.\ Rev.\ D \textbf{104} (2021) 014505 [arXiv:2008.00493 [hep-lat]].

\bibitem{MisraYadav2020}
V.~Yadav and A.~Misra,
\textit{On $M$-theory dual of large-$N$ thermal QCD-like theories up to $\mathcal{O}(R^4)$ and $G$-structure classification of underlying non-supersymmetric geometries},
Adv. Theor. Math. Phys. \textbf{26} (2022) no.10, 3801-3894 [arXiv:2004.07259 [hep-th]].

\bibitem{OnShellR4} The ${\cal O}(R^4)$ on-shell eleven-dimensional action reduces
to two-derivative fluxed gravity evaluated on the ${\cal O}(R^4)$-corrected metric,
\begin{equation*}
\scriptsize
\begin{aligned}
S^{\rm on-shell}
={}&\left.\int d^{11}x\sqrt{-g}
\Big(R-\tfrac{1}{2\cdot4!}G^2+\beta J_0\Big)\right|^{\rm on-shell}
\nonumber\\[-1pt]
={}&\int d^{11}x\Big\{\sqrt{-g[g^{(0)}+\beta g^{(1)}]}
\Big[R[g^{(0)}+\beta g^{(1)}]
\nonumber\\[-1pt]
&\qquad-\tfrac{1}{2\cdot4!}G_4^2(g^{(0)}+\beta g^{(1)})\Big]
+\beta\sqrt{-g^{(0)}}J_0[g^{(0)}]\Big\}^{\rm on-shell}
\nonumber\\[-1pt]
&\qquad+{\cal O}(\beta^2).
\end{aligned}
\end{equation*}
with $J_0[g^{(0)}]\sim{\cal O}(N^{-\gamma})$, $\gamma>1$
\cite{MisraYadav2020}, so the leading ${\cal O}(R^4)$ effect is carried by the
corrected metric $g^{(0)}+\beta g^{(1)}$.

\bibitem{Keshav:inprep} K.~Dasgupta, C.~Gale, S.~Jeon, S.~S.~Kushwah and
A.~Misra, in preparation.

\bibitem{Jain+Samanta+Trivedi} S.~Jain, R.~Samanta and S.~P.~Trivedi,
{\it The Shear Viscosity in Anisotropic Phases}, JHEP \textbf{10}, 028 (2015) [arXiv:1506.01899 [hep-th]].

\bibitem{Demircik2023L}
T.~Demircik, C.~Ecker and M.~J{\"a}rvinen,
Astrophys. J. Lett. \textbf{907} (2021) no.2, L37 [arXiv:2009.10731 [astro-ph.HE]].

\bibitem{Cao2022} Z.~Cao, K.~Hattori, M.~Hongo, X.-G.~Huang and H.~Taya,
\textit{Gyrohydrodynamics: Relativistic spinful fluid with strong vorticity},
PTEP \textbf{2022} (2022) 071D01 [arXiv:2205.08051 [hep-th]].

\bibitem{Symp-half-flat-SU3-struc} A.~Misra, {\it Contact Structure-induced Symplectic Nearly Half-Flat $SU(3)$ Structure in Finite-$N$ Thermal ${\cal M}$-Theory}, arXiv:2608.26267.



\bibitem{JeonYaffe1996} S.~Jeon and L.~G.~Yaffe,
Phys.\ Rev.\ D \textbf{53} (1996) 5799 [hep-ph/9512263].

\bibitem{AMY2003} P.~B.~Arnold, G.~D.~Moore and L.~G.~Yaffe,
JHEP \textbf{01} (2003) 030 [hep-ph/0209353]; JHEP \textbf{05} (2003) 051
[hep-ph/0302165].

\bibitem{lattice-SU3_Glue} H.~B.~Meyer, {\it Calculation of the Bulk Viscosity in SU(3) Gluodynamics}, Phys. Rev. Lett. {\bf 100} (2008) 162001, arXiv:0710.3717 [hep-lat]; H.~B.~Meyer, Phys. Rev. D {\bf 76}, 101701 (2007), arXiv:0704.1801 [hep-lat].

\bibitem{Gerber89} P.~Gerber and H.~Leutwyler,
Nucl.\ Phys.\ B \textbf{321} (1989) 387.

\bibitem{Demircik2023} T.~Demircik, D.~Gallegos, U.~G\"ursoy,
M.~J\"arvinen and R.~Lier,
\textit{A novel method for holographic transport},
[arXiv:2311.00042 [hep-th]].

\bibitem{McTeQ} G.~Yadav, V.~Yadav and A.~Misra,
``${\cal M}$cTEQ (${\cal M}$ chiral perturbation theory-compatible deconfinement Temperature and Entanglement Entropy up to terms Quartic in curvature) and FM (Flavor Memory),'' JHEP \textbf{10}, 220 (2021) [arXiv:2108.05372 [hep-th]].

 







\end{thebibliography}
\end{document}